\documentclass[twocolumn, superscriptaddress,longbibliography]{revtex4-1}
\usepackage{blindtext}
\usepackage{algorithm2e}
\usepackage{float}
\usepackage[utf8]{inputenc}
\usepackage[T1]{fontenc}
\usepackage{amsmath}
\usepackage{subfigure}
\usepackage{ulem}
\usepackage{bm,color,bbm}
\usepackage{hyperref, mathtools}
\usepackage{amsfonts}    
\usepackage{graphicx}   
\usepackage{verbatim}   
\usepackage{booktabs}
\usepackage{threeparttable}
\usepackage{dcolumn}
\usepackage{multirow}
\usepackage{array}
\usepackage{upgreek}
\usepackage{pifont}
\usepackage{textcomp}

\begin{document}
\title{Learning imaging mechanism directly from optical microscopy observations}

\author{Ze-Hao Wang}
\email{These authors contribute equally.}
\affiliation{CAS Key Laboratory of Quantum Information, University of Science and Technology of China, Hefei, 230026, China}
\affiliation{CAS Center For Excellence in Quantum Information and Quantum Physics, University of Science and Technology of China, Hefei, 230026, China}

\author{Long-Kun Shan}
\email{These authors contribute equally.}
\affiliation{CAS Key Laboratory of Quantum Information, University of Science and Technology of China, Hefei, 230026, China}
\affiliation{CAS Center For Excellence in Quantum Information and Quantum Physics, University of Science and Technology of China, Hefei, 230026, China}

\author{Tong-Tian Weng}
\affiliation{CAS Key Laboratory of Quantum Information, University of Science and Technology of China, Hefei, 230026, China}
\affiliation{CAS Center For Excellence in Quantum Information and Quantum Physics, University of Science and Technology of China, Hefei, 230026, China}

\author{Tian-Long Chen}
\affiliation{University of Texas at Austin, Austin, TX 78705, USA}

\author{Qi-Yu Wang}
\affiliation{CAS Key Laboratory of Quantum Information, University of Science and Technology of China, Hefei, 230026, China}
\affiliation{CAS Center For Excellence in Quantum Information and Quantum Physics, University of Science and Technology of China, Hefei, 230026, China}

\author{Xiang-Dong Chen}
\affiliation{CAS Key Laboratory of Quantum Information, University of Science and Technology of China, Hefei, 230026, China}
\affiliation{CAS Center For Excellence in Quantum Information and Quantum Physics, University of Science and Technology of China, Hefei, 230026, China}
\affiliation{Hefei National Laboratory, University of Science and Technology of China, Hefei 230088, China}

\author{Zhang-Yang Wang}
\affiliation{University of Texas at Austin, Austin, TX 78705, USA}

\author{Guang-Can Guo}
\affiliation{CAS Key Laboratory of Quantum Information, University of Science and Technology of China, Hefei, 230026, China}
\affiliation{CAS Center For Excellence in Quantum Information and Quantum Physics, University of Science and Technology of China, Hefei, 230026, China}
\affiliation{Hefei National Laboratory, University of Science and Technology of China, Hefei 230088, China}

\author{Fang-Wen Sun}
\email{fwsun@ustc.edu.cn}
\affiliation{CAS Key Laboratory of Quantum Information, University of Science and Technology of China, Hefei, 230026, China}
\affiliation{CAS Center For Excellence in Quantum Information and Quantum Physics, University of Science and Technology of China, Hefei, 230026, China}
\affiliation{Hefei National Laboratory, University of Science and Technology of China, Hefei 230088, China}

\date{\today}

\begin{abstract}
    Optical microscopy image plays an important role in scientific research through the direct visualization of the nanoworld, where the imaging mechanism is described as the convolution of the point spread function (PSF) and emitters. Based on a priori knowledge of the PSF or equivalent PSF, it is possible to achieve more precise exploration of the nanoworld. However, it is an outstanding challenge to directly extract the PSF from microscopy images. Here, with the help of self-supervised learning, we propose a physics-informed masked autoencoder (PiMAE) that enables a learnable estimation of the PSF and emitters directly from the raw microscopy images. We demonstrate our method in synthetic data and real-world experiments with significant accuracy and noise robustness. PiMAE outperforms DeepSTORM and the Richardson-Lucy algorithm in synthetic data tasks with an average improvement of 19.6\% and 50.7\% (35 tasks), respectively, as measured by the normalized root mean square error (NRMSE) metric. This is achieved without prior knowledge of the PSF, in contrast to the supervised approach used by DeepSTORM and the known PSF assumption in the Richardson-Lucy algorithm. Our method, PiMAE, provides a feasible scheme for achieving the hidden imaging mechanism in optical microscopy and has the potential to learn hidden mechanisms in many more systems.

\end{abstract}


\flushbottom
\maketitle
\thispagestyle{empty}
\hfill

\section{Introduction}
    Optical microscopy is of great importance in scientific research to observe the nanoworld. The common view is that the Abbe diffraction limit describes the lower bound of the spot size and thus limits the microscopic resolution. However, recent studies have demonstrated that by designing and measuring the PSF or equivalent PSF of microscopy, it is possible to achieve sub-diffraction-limit localization of emitters. Techniques such as photoactivated localization microscopy \cite{lee2012counting} and stochastic optical reconstruction microscopy \cite{rust2006sub} attain super-resolution molecular localization through selective excitation and reconstruction algorithms that are based on the microscopy PSF. Spatial mode sorting-based microscopic imaging method (SPADE  \cite{bearne2021confocal}) can be treated as a deconvolution problem using higher-order modes as the equivalent PSF. Stimulated-emission depletion microscopy achieves super-resolution imaging by introducing illumination with donut-shaped PSFs to selectively deactivate fluorophores \cite{hell1994breaking, chen2015subdiffraction}. Additionally, deep learning-based methods, such as DeepSTORM \cite{nehme2018deep} and DECODE \cite{speiser2021deep}, use deep neural networks (DNNs) to predict emitters in raw images by synthesizing training sets with the same PSFs as used in actual experiments. In all of these microscopic imaging techniques, a prior knowledge of the PSF is crucial, making it of great interest to develop a method for directly estimating the PSF from raw images.

    Currently, some traditional algorithms such as Deconvblind \cite{biggs1997acceleration} use maximum likelihood estimation to infer the PSF and emitters from raw images \cite{chan1998total, krishnan2011blind, liu2014blind, michaeli2014blind, pan2017deblurring, pan2016l_0, ren2016image, sun2013edge, yan2017image, zuo2016learning}. However, these algorithms face two challenges. Firstly, they struggle to estimate PSFs with complex shapes. Secondly, they can lead to trivial solutions where the PSF is a $\delta$-function and the image of the emitters is equal to the raw image. To tackle these issues, researchers have turned to using Deep Neural Networks (DNNs) \cite{shajkofci2020spatially}. However, this requires a library of PSFs and a large amount of sharp microscope images to generate the training dataset, which limits the application of these algorithms.

    \begin{figure*}[htbp]
        \centering
        \includegraphics[width=0.8\linewidth]{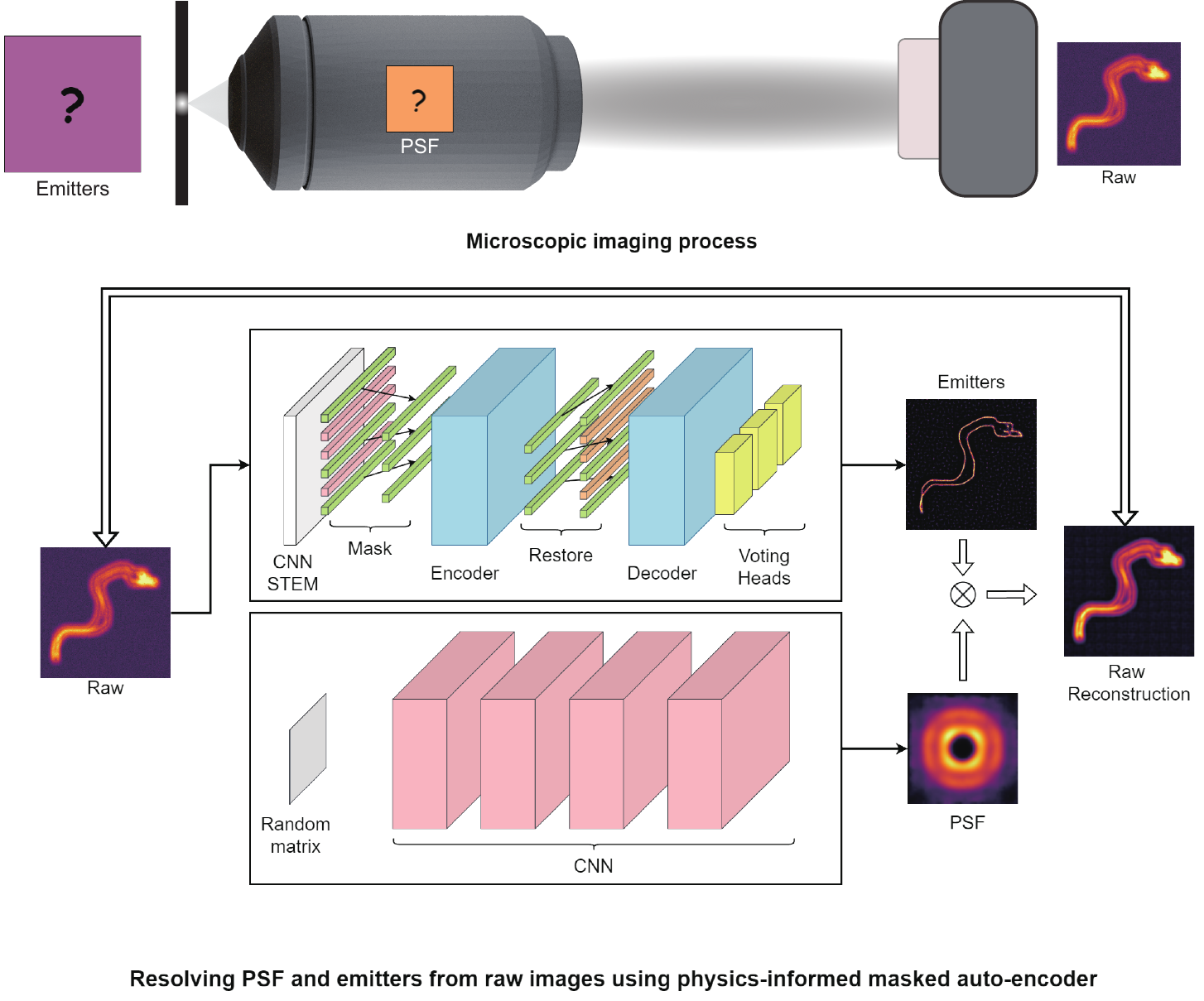}
        \caption{\textbf{PiMAE Overview.} PiMAE, a physics-informed masked autoencoder, is proposed to learn the imaging mechanism of an optical microscope.}
        \label{fig:PiMAE}
    \end{figure*}
    We use self-supervised learning to overcome the above challenges. Here, we treat the PSF as the pattern hidden in the raw images and the emitters as the sparse representation of the raw image. As a result, we propose a physics-informed masked autoencoder (PiMAE, Fig \ref{fig:PiMAE}) that estimates the PSF and emitters directly from the microscopy raw images. Using raw data synthesized by various simulated PSFs, we compare the results of PiMAE and Deconvblind \cite{biggs1997acceleration} for estimating PSF, as well as PiAME, Richardson-Lucy algorithm \cite{lucy1974iterative} and DeepSTORM \cite{nehme2018deep} for localizing emitters. Our proposed self-supervised learning approach, PiMAE, outperforms existing algorithms without the need for data annotation or PSF measurement. PiMAE demonstrates a significant performance improvement, as measured by the NRMSE metric, and is highly resistant to noise. In tests with real-world experiments, PiMAE resolves wide-field microscopy images of standard PSF, out-of-focus PSF, and aberrated PSF with high quality, and the results achieve a resolution comparable to SIM results. Also, we demonstrates that 5 raw images can satisfy the requirements of self-supervised training. This approach, PiMAE, shows wide applicability in synthetic data testing and real-world experiments. We expect its usage for estimation of hidden mechanisms in various physical systems.

\section{Method}
    Self-supervised learning leverages the inherent structure or patterns in data to learn meaningful representations. There are two main categories: Contrastive Learning \cite{oord2018representation, wu2018unsupervised, he2020momentum, chen2020simple} and pretext task learning \cite{doersch2015unsupervised, dosovitskiy2014discriminative, devlin2018bert, chen2020adversarial, chen2020self}. Mask Image Modeling (MIM) \cite{chen2020generative, doersch2015unsupervised, henaff2020data, pathak2016context, trinh2019selfie} is a pretext task learning technique that randomly masks portions of an input image. Recently, MIM has been shown to learn transferable, robust, and generalized representations from visual images, improving performance in downstream computer vision tasks \cite{he2021masked}. PiMAE is a MIM-based method that reconstructs raw images according to the imaging principle of optical microscopy.

    \subsection{PiMAE model.}
        The PiMAE model (Figure \ref{fig:PiMAE}) consists of three key components: (1) a Vision Transformer-based \cite{dosovitskiy2020image} encoder-decoder architecture with a mask layer to prevent trivial solutions while estimating emitters, (2) a Convolutional Neural Network as a prior for PSF estimation \cite{ulyanov2018deep}, and (3) a microscopic imaging process that enforces adherence to the microscopy principle. Appendix \ref{nn_arch} provides detailed information on the network architecture and the embedding of physical principle. PiMAE requires only a few raw images for training, which is attributed to the carefully designed loss function. The loss function consists of two parts: one measures the difference between the raw and the reconstruction images, including the mean of the absolute difference and the multi-scale structure similarity; the other part is a constraint on the PSF, including the total variation loss measuring the PSF continuity and the offset distance of the PSF's center of mass. Appendix \ref{loss} contains the expressions for the loss functions.

        \begin{figure*}[htbp]
            \centering
            \includegraphics[width=0.9\linewidth]{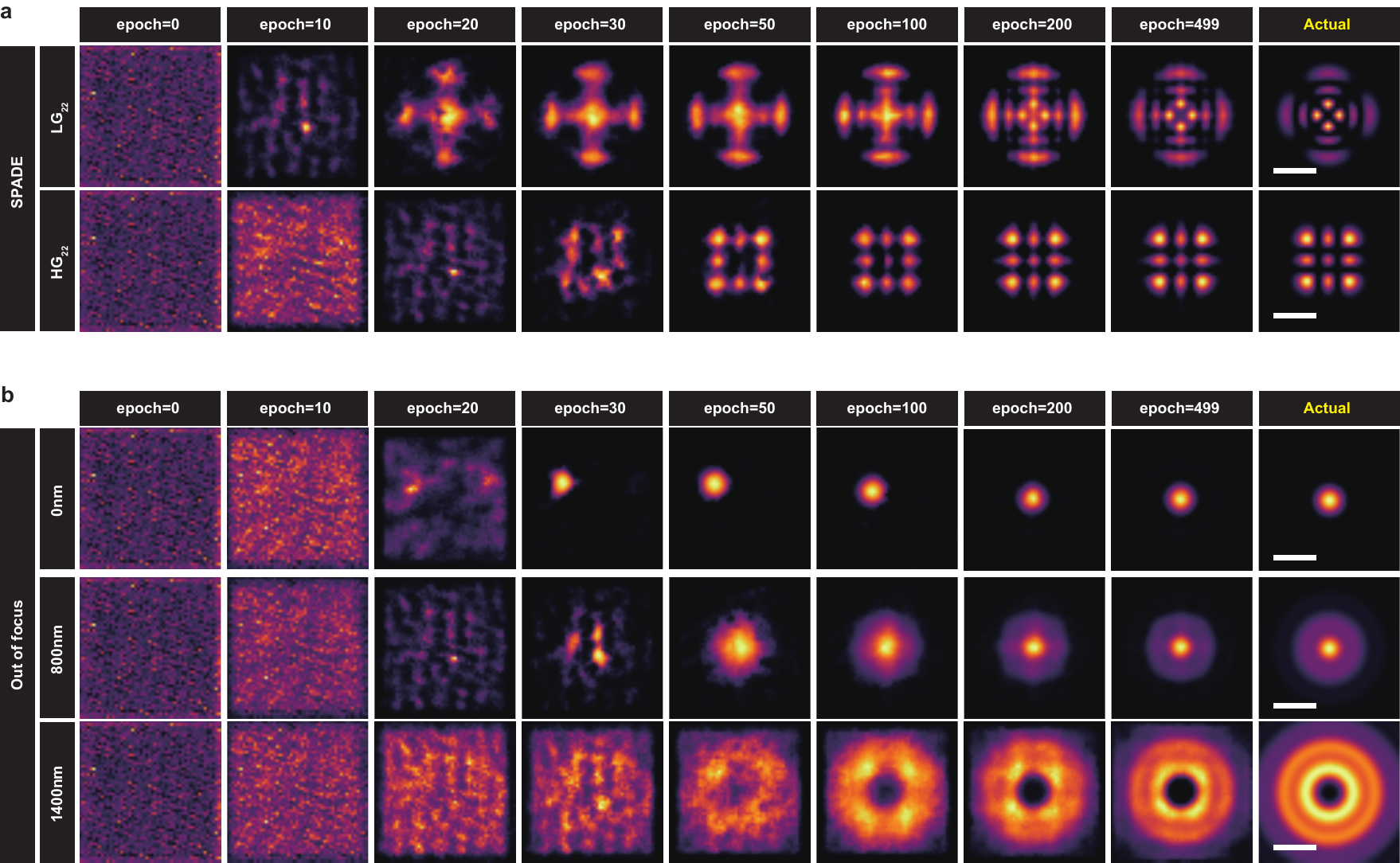}
            \caption{\textbf{PSF learning.} The scale bar is 0.5 \textmu m. The results demonstrate that PiMAE can successfully learn the PSF from raw images through the training process. \textbf{a} The figure displays the PSF of SPADE, include Laguerre-Gaussian mode $\text{LG}_{\text{22}}$ and Hermite-Gaussian mode $\text{HG}_{\text{22}}$. \textbf{b} different out-of-focus distances (800 nm and 1400 nm) under a wide-field microscope imaging setup, along with the in-focus image.}
            \label{fig:psf}
        \end{figure*}

        \begin{figure*}[htbp]
            \centering
            \includegraphics[width=0.78\linewidth]{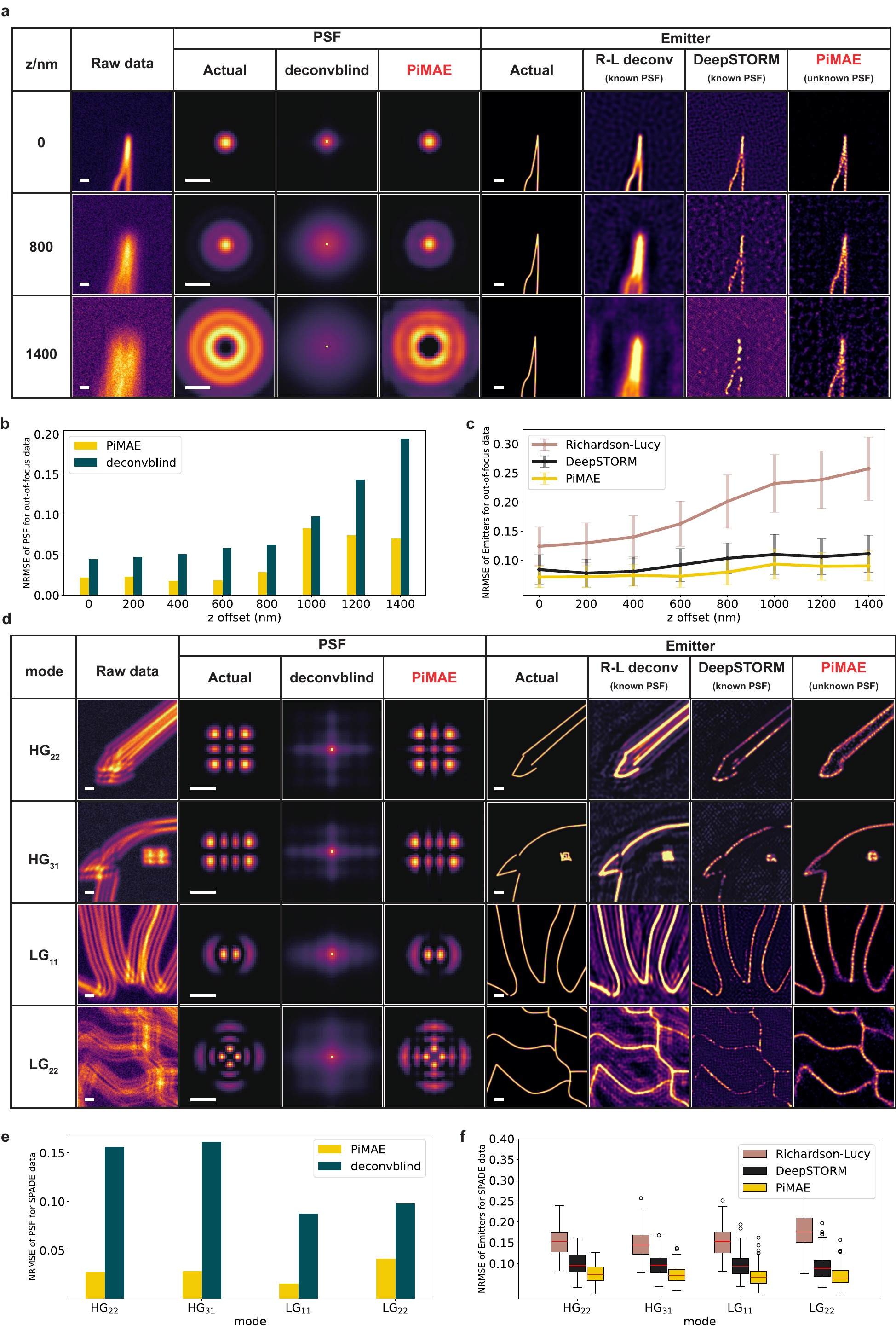}
            \caption{\textbf{Evaluation in synthetic datasets.}  The scale bar is 0.5 \textmu m. \textbf{a.} The results of estimated PSF and emitters from out-of-focus synthetic data. \textbf{b.} NRMSE of the results of estimated PSF from out-of-focus synthetic data. \textbf{c.} NRMSE of the results of estimated emitters from out-of-focus synthetic data. \textbf{d.} The results of estimated PSF and emitters from synthetic data with Hermite-Gaussian mode and Laguerre-Gaussian mode (HG/LG) as PSF. \textbf{e.} NRMSE of the results of estimated PSF from HG/LG synthetic data. \textbf{f.} NRMSE of the results of estimated emitters from HG/LG synthetic data. The noise scale in the above evaluations is $ \text{noise}_{\text{std}}/\text{raw}_{\text{mean}}=0.5 $.}
            \label{fig:synthetic}
        \end{figure*}
    \subsection{Training.}
        The Vision Transformer (ViT) based encoder in PiMAE is pre-trained on the COCO dataset \cite{lin2014microsoft} to improve performance. This pre-training relies on the self-supervised learning of a masked autoencoder, but does not incorporate any physical information (detailed in Appendix \ref{Pretrain}). After pre-training, PiMAE loads the trained encoder parameters and undergoes self-supervised training using raw microscopic images. The input image size is 144 pixels, and we use the RAdam optimizer \cite{liu2019radam} with a learning rate of $1e^{-4}$ and a batch size of 18. The training runs for $5e^4$ steps.

        Within PiMAE, the convolutional neural network, depicted in Figure \ref{fig:PiMAE}, is initialized randomly and takes a fixed random vector as input, outputting the predicted PSF. Relevant details can be found in Appendix \ref{nn_arch}. As PiMAE undergoes self-supervised training, the CNN's predicted PSF continually becomes more accurate, moving closer to the true PSF as shown in Figure \ref{fig:psf}. The experimental setup is shown in Figure \ref{fig:synthetic}.

    \subsection{Synthetic data design}
        To evaluate PiMAE's performance, synthetic datasets were designed taking into account the following factors: (1) PiMAE's requirement for sparse emitter data, (2) the need for the emitter data to avoid discrete points for more challenging PSF estimation tasks, (3) the inclusion of both standard Gaussian shapes and other challenging PSF shapes, (4) raw image noise, and (5) emitter sparsity. The Sketches dataset \cite{eitz2012hdhso} was chosen as the emitter, as described in Appendix \ref{Sketches}, and various commonly used PSFs were designed in Appendix \ref{PSFs}. The robustness to noise was evaluated by adding noise to the raw images at different levels. Moreover, images with sparse lines of varying densities were generated as emitters to assess the impact of sparsity on PiMAE, as described in Appendix \ref{Randomlines}.

    \subsection{Real-world experiments}
        We evaluate PiMAE's performance in handling both standard and non-standard PSF microscopy images in real-world experiments. Since the true emitter positions cannot be obtained, we use the BioSR \cite{qiao2021evaluation} dataset to evaluate PiMAE's handling of standard PSF microscopy images and compare it with structured illumination microscopy (SIM). Then, we use our custom-made wide-field microscope to produce out-of-focus and distorted PSF microscopy images to qualitatively analyze PiMAE's performance in handling non-standard PSF microscopy images.
        \subsubsection{Wide field microscopic imaging of NV color centers}
            A 532 nm (COHERENT Vendi 10 Single longitudinal mode laser) laser passes through a customized precision electronic timing shutter, which controls the duration of the laser beams flexibly. The laser is then expanded and sent to polarization mode controller which consists of a polarizing film (LPVISE100-A) and a half wave plate (thorlabs WPH10ME-532). The extended laser is focused to the focal plane behind the objective lens (Olympus, UPLFLN100XO2PH) by a fused quartz lens with a focal length of 150 mm. The fluorescence signals are collected by a scientific complementary metal oxide semiconductor (sCMOS) camera (Hamamatsu, Orca Flash 4.0 v.3). We use a manual zoom lens (NIKON AF 70-300 mm,f/4-5.6G, focal length between 70 mm and 300 mm, and the field of view of 6.3) as tubelens to continuously change the magnification of the microscopic system.

\section{Result}
    \subsection{PiMAE achieves high accuracy on synthetic datasets.}
        We use samples from the Sketches \cite{eitz2012hdhso} dataset as emitters and synthesize raw data for various microscopy scenarios by convolving the emitters with the simulated PSFs. For each scenario, we sample 1000 images from the Sketches dataset as the training set and 100 images as the test set. In Appendix \ref{training_set_size}, we show that PiMAE can achieve good training results even with a minimum of 5 images in the training set. To evaluate the PSF estimasution, we use Deconvblind \cite{biggs1997acceleration} as benchmark. To evaluate the localization accuracy of emitters, we utilize the Richardson-Lucy algorithm \cite{lucy1974iterative} and DeepSTORM \cite{nehme2018deep} as reference methods. The results are measured by NRMSE. It is worth noting that the PSF is assumed to be known a priori during the testing of the Richardson-Lucy algorithm and DeepSTORM, while in the case of PiMAE, the PSF is treated as unknown.

        Out-of-focus is one of the most common factors that can affect the quality of microscope imaging. PiMAE is capable of addressing this issue, and we demonstrate this by simulating a range of wide field microscopy PSFs with out-of-focus distances that vary from 0 nm to 1400 nm. We also add Gaussian noise with a scale of $ \text{noise}_{\text{std}}/\text{raw}_{\text{mean}}=0.5 $ to raw images, where $\text{noise}_{\text{std}}$ is the standard deviation of Gaussian noise \cite{fisher2013image} and $\text{raw}_{\text{mean}}$ is the mean value of the raw image. First, we evaluate the performance of estimated PSFs. Figure \ref{fig:synthetic}a shows the actual PSFs and those estimated by Deconvblind and PiMAE. The PiMAE estimated PSF is similar to the actual PSF for all out-of-focus distances, while most of Deconvblind's estimated PSFs are far from the truth, indicating that Deconvblind cannot resolve raw images with complex PSFs. Furthermore, the estimated PSF by Deconvblind converges to the $\delta$-function after several iterations (see Appendix \ref{trival}). The NRMSE of the estimated PSFs at different out-of-focus distances is quantified in Figure \ref{fig:synthetic}b, with PiMAE achieving much better results than Deconvblind. Second, we evaluate the performance of estimated emitters. Figure \ref{fig:synthetic}a also shows the actual emitters and those estimated by the Richardson-Lucy algorithm, DeepSTORM, and PiMAE. When the out-of-focus distance is large, PiMAE and DeepSTORM significantly outperform the Richardson-Lucy algorithm. The NRMSE at different out-of-focus distances is shown in Figure \ref{fig:synthetic}c, with PiMAE achieving the best performance despite having no knowledge of the actual PSF.
    
        Recently, researchers have found that imaging resolution can be improved using a spatial pattern sorter \cite{tsang2016quantum, bearne2021confocal, shajkofci2020spatially}, a method referred to as SPADE. Using SPADE for confocal microscopy is equivalent to using PSFs corresponding to spatial modes \cite{bearne2021confocal}, such as Zernike modes, Hermite-Gaussian (HG) modes an\textbf{}d Laguerre-Gaussian (LG) modes. However, SPADE faces several challenges, including the need for an accurate determination of the spatial mode (i.e., the equivalent PSF), sensitivity to noise, and a lack of reconstruction algorithms for complex spatial modes. PiMAE can solve these problems.  Figure \ref{fig:synthetic}(d-f) show the SPADE imaging results with noise scale $ \text{noise}_{\text{std}}/\text{raw}_{\text{mean}}=0.5 $. PiMAE can accurately estimate the equivalent PSF and emitters, and the performance is much better than that of the Deconvblind, Richardson-Lucy algorithm and DeepSTORM. Therefore, PiMAE can significantly improve the performance of SPADE. These experiments demonstrate that PiMAE is effective for scenarios with unknown and complex imaging PSFs.

    \begin{figure}[htbp]
        \centering
        \includegraphics[width=0.70\linewidth]{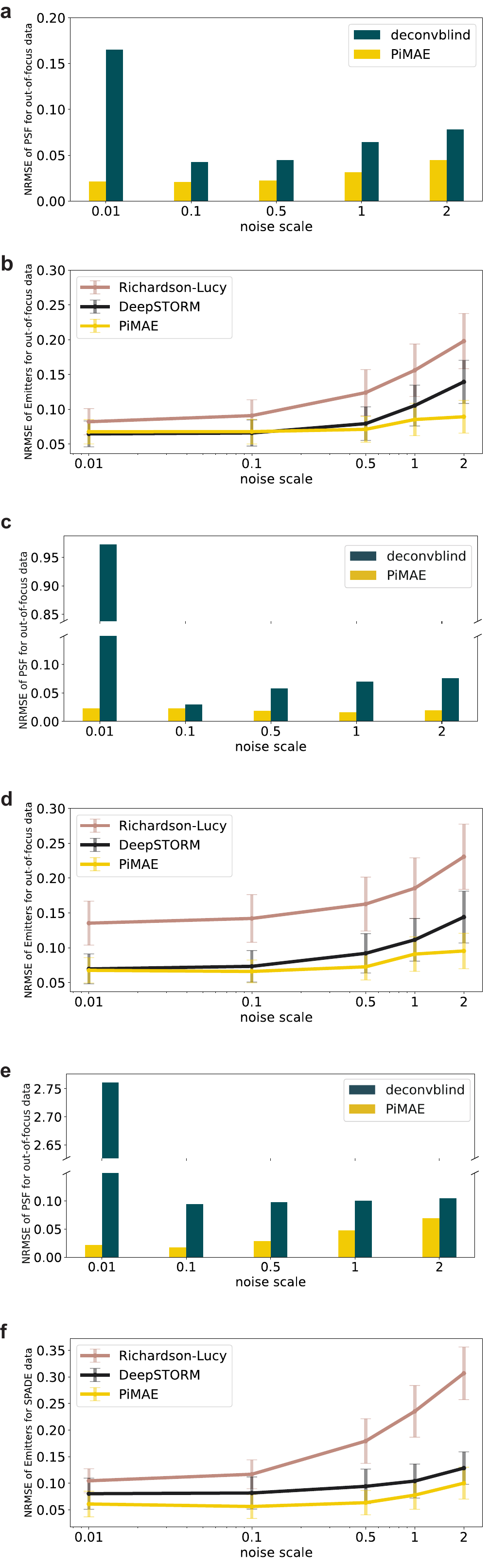}
        \caption{\textbf{Evaluation of noise robustness.} \textbf{a.} NRMSE of the results of estimated PSF from in-focus synthetic data. \textbf{b.} NRMSE of the results of estimated emitters from in-focus synthetic data. \textbf{c.} NRMSE of the results of estimated PSF from 600 nm out-of-focus synthetic data. \textbf{d.} NRMSE of the results of estimated emitters from 600 nm out-of-focus synthetic data. \textbf{e.} NRMSE of the results of estimated PSF from $\text{LG}_{\text{22}}$ synthetic data. \textbf{f.} NRMSE of the results of estimated emitters from $\text{LG}_{\text{22}}$ synthetic data. The noise scale is $ \text{noise}_{\text{std}}/\text{raw}_{\text{mean}}$. }
        \label{fig:noise}
    \end{figure}

    \subsection{Noise robustness}
        Noise robustness is a crucial metric for evaluating reconstruction algorithms. We evaluate noise robustness in three scenarios: (1) in-focus wide field microscopy; (2) wide-field microscopy at 600 nm out-of-focus distance; (3) Laguerre-Gaussian mode $\text{LG}{22}$ SPADE imaging. The raw image of each scenario contains Gaussian noise at scales ($ \text{noise}_{\text{std}}/\text{raw}_{\text{mean}}$) of 0.01, 0.1, 0.5, 1, and 2, as shown in Figure \ref{fig:noise}. We first compare the results of Deconvblind and PiMAE for estimating PSF. We find that PiMAE shows excellent noise immunity, substantially outperforming Deconvblind in all tests. We then compare the results of the Richardson-Lucy algorithm, DeepSTORM, and PiMAE for estimating the emitters. Overall, PiMAE performs the best, only slightly behind DeepSTORM in the standard PSF results at low noise. The Richardson-Lucy algorithm performs similarly to DeepSTORM and PiMAE when the noise scale is very small. However, when the noise scale slightly increases, its performance significantly decreases. This shows the advantage of deep learning-based methods over traditional algorithms in terms of noise robustness. Moreover, the advantage of PiMAE over the other two algorithms increases as the scale of the noise becomes larger and the shape of the PSF becomes more complex.

    \begin{figure}[htbp]
        \centering
        \includegraphics[width=\linewidth]{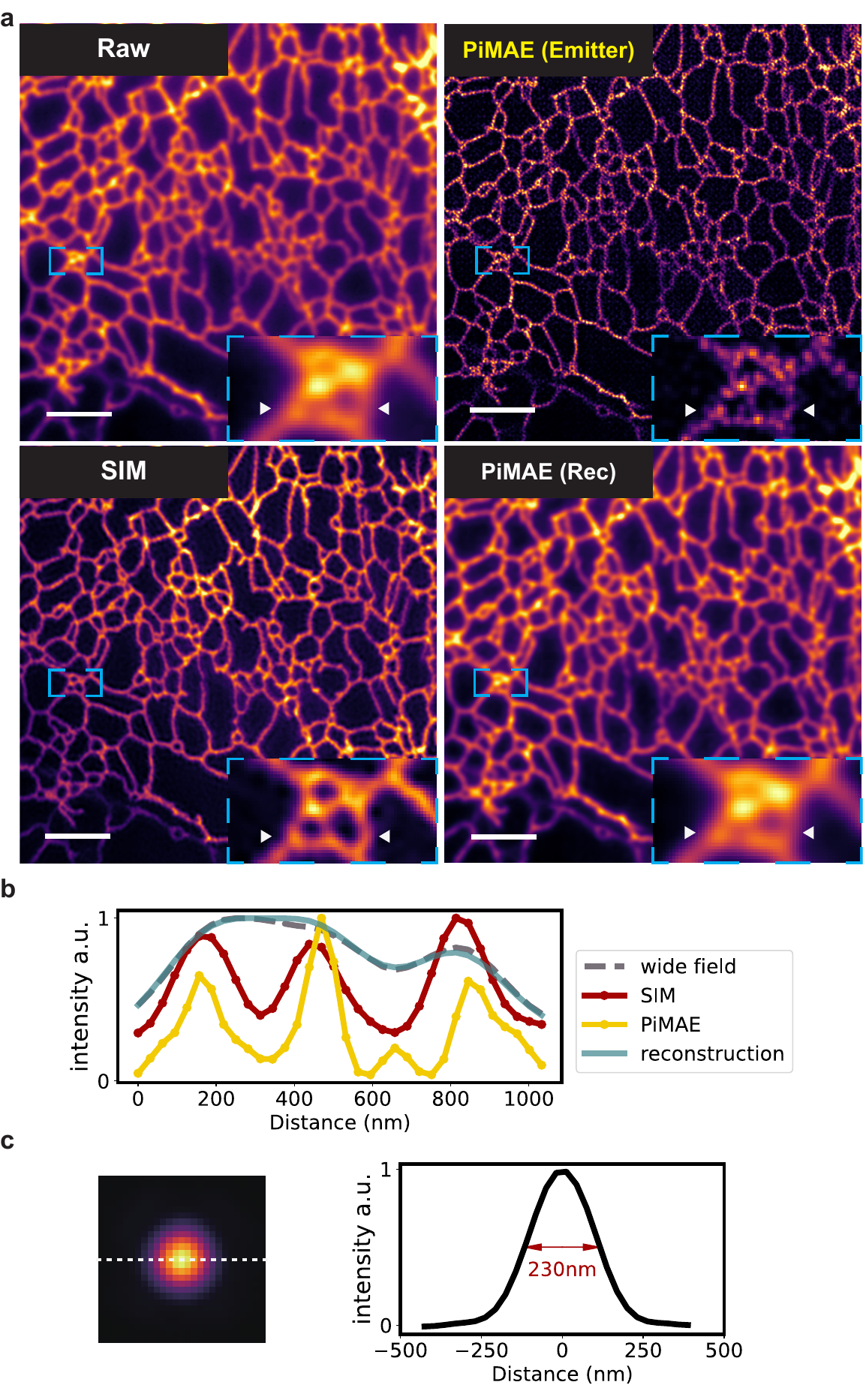}
        \caption{\textbf{Super-resolution imaging of ER.} The scale bar is 2.50 \textmu m. \textbf{a.} The figures are the raw image of wide-field microscopic imaging of ER, the result of estimating the emitter from wide-field microscopic imaging using PiMAE, the result of SIM of the same field of view and the result of wide-field microscopic imaging reconstructed by PiMAE. Data from BioSR dataset \cite{qiao2021evaluation}. \textbf{b.} Comparison of the cross section of the PiMAE estimated emitters and SIM results. It shows that the resolution of the results obtained by PiMAE is comparable to that of SIM. \textbf{c.}  PiMAE estimated wide-field microscope PSF with a FWHM of 230 nm, where the diffraction limit is \text{229 nm}.}
        \label{fig:biosr_er}
    \end{figure}

    \subsection{PiMAE enables image super-resolution for wide-field microscopy comparable to SIM}
        The endoplasmic reticulum (ER) is a system of tunnels surrounded by membranes in eukaryotic cells. In the dataset BioSR \cite{qiao2021evaluation}, the researchers imaged the ER in the same field of view using wide-field microscopy and SIM, respectively. Figure \ref{fig:biosr_er}a shows the results of PiMAE-resolved wide-field microscopy raw images. We find that the resolution of the PiMAE estimated emitter is comparable to that of SIM, which has a resolution twice that of the diffraction limit. Figure \ref{fig:biosr_er}b shows the cross section results, where the peak positions of the PiMAE estimated emitter match the peak positions of the SIM results, corresponding to indistinguishable wide-field imaging results. This indicates that the resolvability of the results of wide-field microscopy with PiMAE estimated emitters is improved to a level similar to that of SIM. Figure \ref{fig:biosr_er}c shows the results of the PiMAE estimated PSF with FWHM of 230 nm. The fluorescence wavelength of the raw image is 488 nm, the numerical aperture (NA) is 1.3, and its diffraction limit is $ 0.61 \times \frac{\lambda}{\text{NA}} = 0.61\times \frac{\text{488 nm}}{1.3}\approx\text{229 nm} $, which is very close to the FWHM of the PiMAE estimated PSF. This experiment shows that PiMAE can be applied to real-world experiments to estimate PSF from raw microscopy data and further improve resolution.

    \subsection{PiMAE enables imaging for non-standard wide-field microscopy}
\begin{figure*}[htbp]
            \centering
            \includegraphics[width=0.75\linewidth]{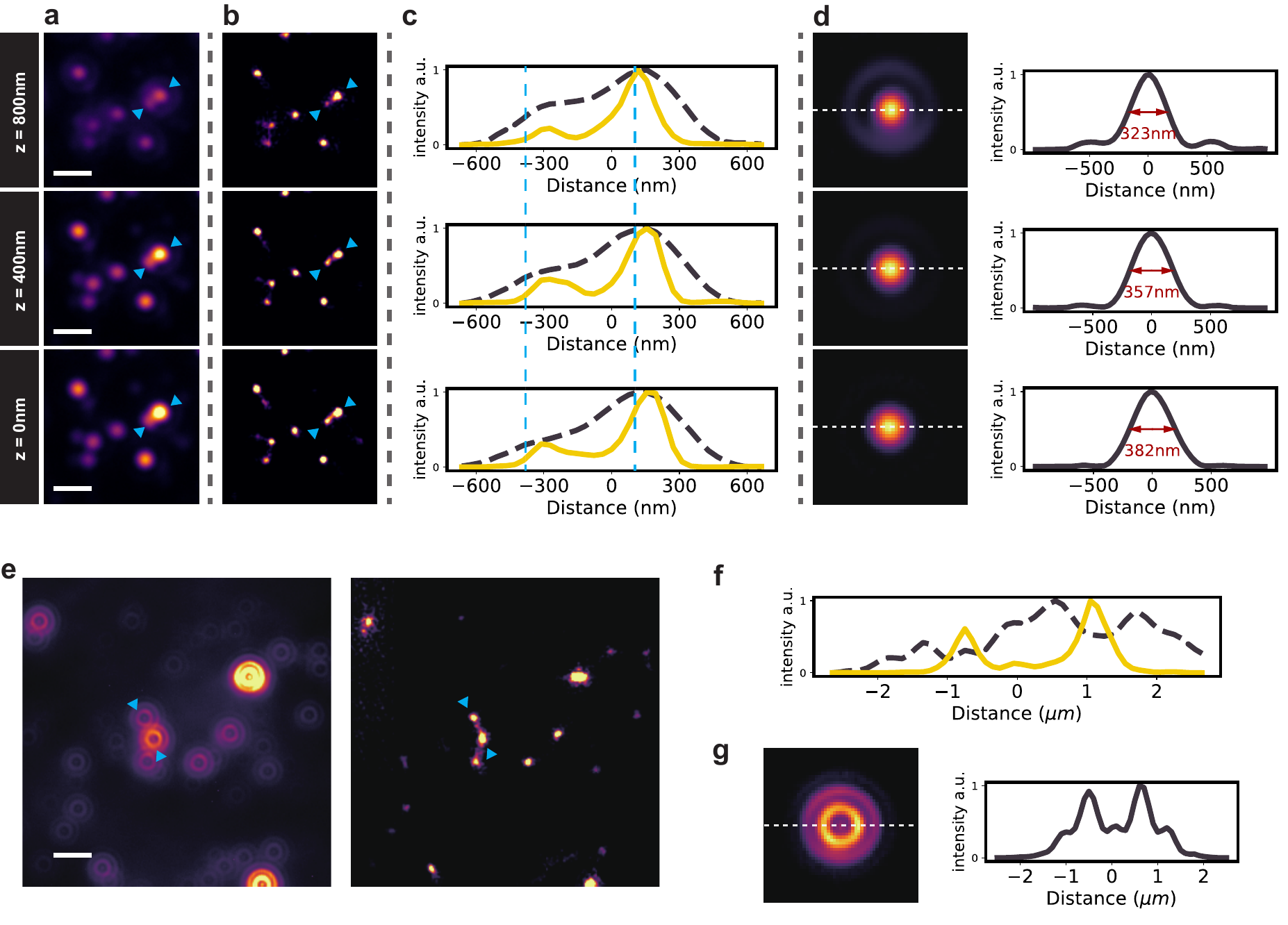}
            \caption{\textbf{Wide-field microscopy imaging of NV color centers.}  \textbf{a-d.} Results of wide-field microscopy imaging of NV color centers at different out-of-focus distances. \textbf{a.} Raw images. The length of the scale bar is 1.25 \textmu m. \textbf{b.} PiMAE estimated emitters. \textbf{c.} The comparison of the cross section of the raw images and the PiMAE-estimated emitters, where the black dashed line represents the raw images and the yellow solid line represents the PiMAE-estimated emitters. The peak positions of the PiMAE-estimated emitter results are constant for different out-of-focus distances, as seen from the blue dashed line. \textbf{d.} The PiMAE estimated PSF.  The FWHM of in-focus PSF is 382 nm, where the diffraction limit is $\text{384 nm} $. The larger the out-of-focus distance, the larger the paraflap of the PSF, despite the decrease of the FWHM in the center region. \textbf{e.} The comparison of non-standard microscopic imaging and PiMAE estimated emitters in the case of mismatched objective and coverslip. The length of the scale bar is 3.2 \textmu m. \textbf{f.} The cross section of the non-standard microscopic imaging and PiMAE estimated emitters. \textbf{g.} The PiMAE-estimated non-standard microscopy PSF. }
            \label{fig:res_nv_a}
        \end{figure*}

        The nitrogen-vacancy (NV) color center is a point defect in diamond which is widely used in super-resolution microscopy \cite{chen2015subdiffraction, han2009three} and quantum sensing \cite{chen2021focusing, degen2017quantum}. We make a home-built wide field microscope to image NV center in fluorescent nanodiamonds (FND) at out-of-focus distances of 0 nm, 400 nm and 800 nm. We take 10 raw images with a size of 2048 pixels and a field of view size of 81.92 \textmu m at each out-of-focus distances. Figure \ref{fig:res_nv_a}a shows that we image NV color centers in the same field of view at different out-of-focus distance and Figure \ref{fig:res_nv_a}b shows the corresponding PiMAE estimated emitters. This is a side-by-side demonstration of the accuracy of the PiMAE estimated emitters. The out-of-focus distance changes during the experiment, but the field of view is invariant. Therefore, the PiMAE estimated emitter position should be constant at each out-of-focus distance, as we observe in Figure \ref{fig:res_nv_a}b-c. Figure \ref{fig:res_nv_a}d shows that the variation of the PSF. The asymmetry of the PSF comes from the slight tilt of the carrier stage. Also we show the PSF cross section for each scene. The FWHM of the estimated PSF at focus is 382 nm, which corresponds to a diffraction limit of 384 nm. This suggests that PiMAE can be applied in real-world experiments to improve the imaging capabilities of microscopes suffering from out-of-focus.

        Moreover, we construct a non-standard PSF for wide-field microscopic imaging of NV color centers by making the objective mismatch with the coverslip, and the results are shown in Figure \ref{fig:res_nv_a}e-g. Figure \ref{fig:res_nv_a}e shows the imaging results and PiMAE estimated emitters. Figure \ref{fig:res_nv_a}f shows the results of the cross-sectional comparison. Figure \ref{fig:res_nv_a}g shows the PiMAE estimated PSF. This experiment demonstrates that PiMAE enables researchers to use microscopy with non-standard PSFs for imaging. And PiMAE's ability to resolve non-standard PSFs expands the application scenarios of NV color centers in fields such as quantum sensing and bioimaging.

    \subsection{PiMAE enables microscopy imaging with widely spread out PSF}
\begin{figure}[htbp]
            \centering
            \includegraphics[width=\linewidth]{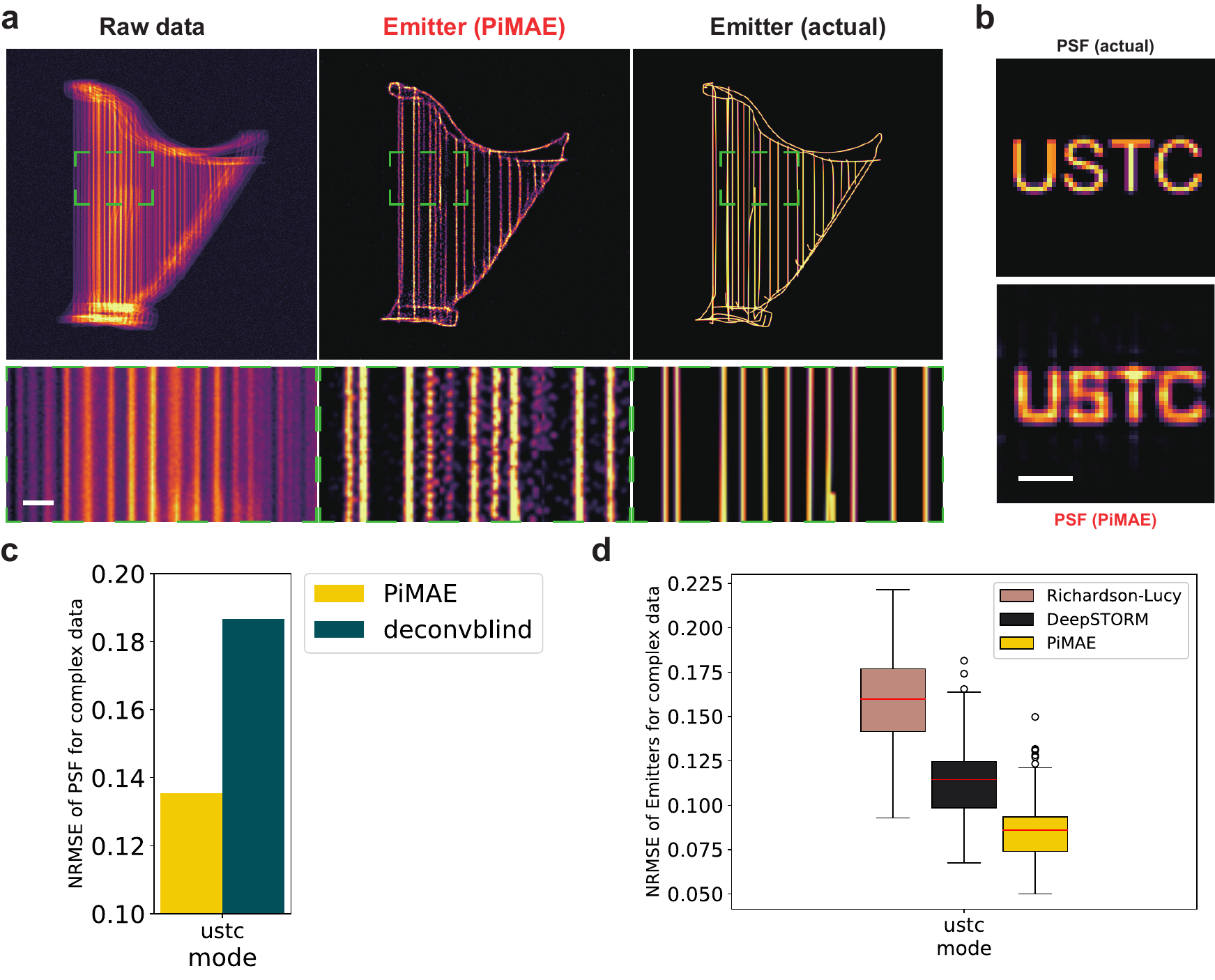}
            \caption{\textbf{Evaluation using synthetic data based on PSF of the shape "USTC".} The scale bar is 0.5 \textmu m. \textbf{a.} The comparison of the raw image, the PiMAE estimated emitters and the actual emitters. \textbf{b.} The comparison of the actual PSF and the PiMAE-estimated PSF. \textbf{c.} NRMSE of the estimated PSF.  \textbf{d.} NRMSE of the estimated emitters.}
            \label{fig:res_ustc_a}
        \end{figure}

        Further testing the capabilities of PiMAE, we evaluate the performance of PiMAE on complex widely spread out PSF, represented by the character "USTC". We use 1000 images as the training set and 100 images as the test set. The noise level is set at $\text{noise}_{\text{std}}/\text{raw}_{\text{mean}}=0.01$. The results of the raw images, the PiMAE processed images, and the evaluation of the NRMSE metric are depicted in Figure \ref{fig:res_ustc_a}.  PiMAE performed exceptionally well, demonstrating its effectiveness in difficult scenarios.
        
    \subsection{Evaluation of the influence of emitter sparsity}

\begin{figure*}[htbp]
            \centering
            \includegraphics[width=0.75\linewidth]{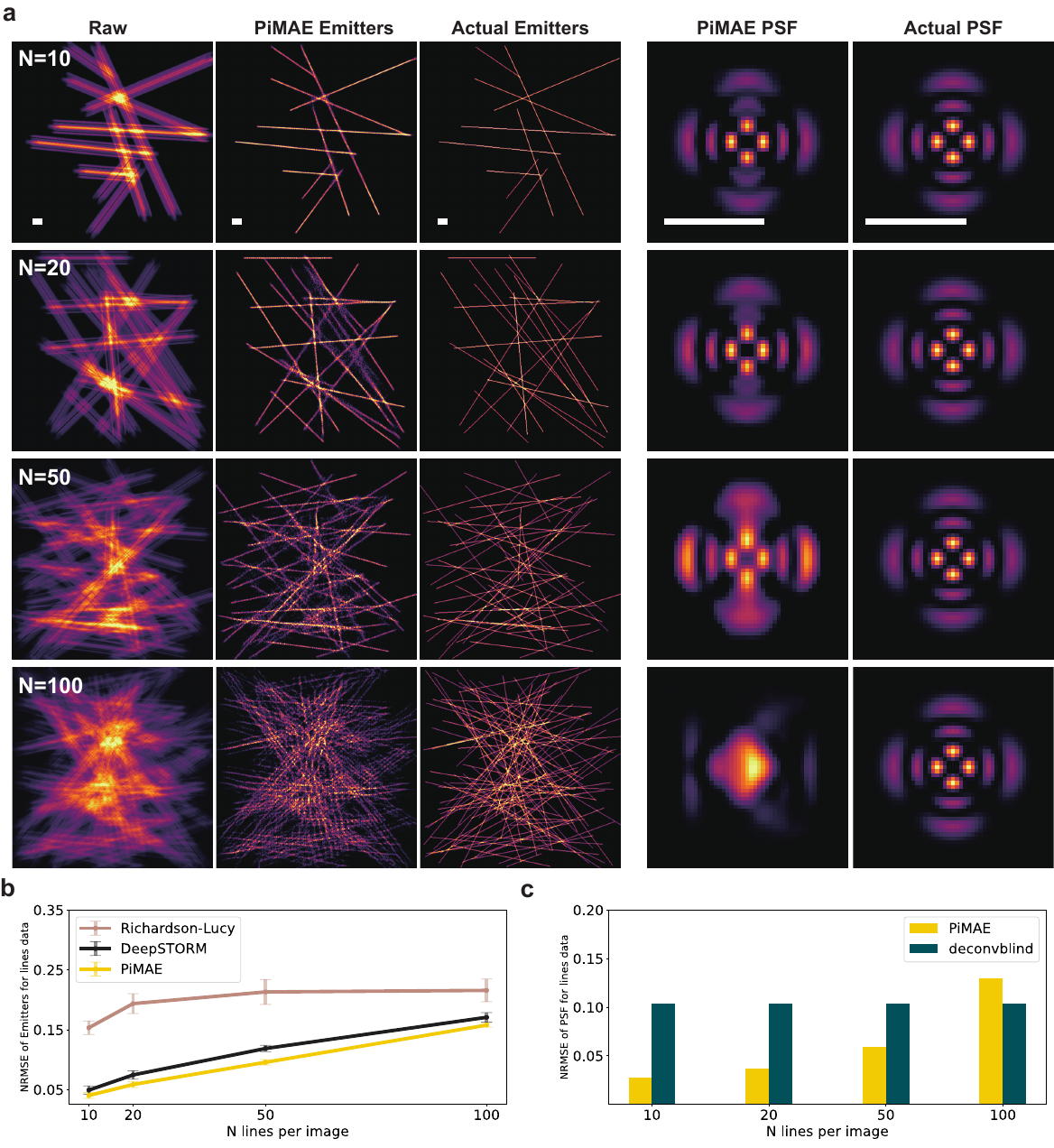}
            \caption{\textbf{The influence of emitter sparsity} The scale bar is 1.0 \textmu m. \textbf{a.} The comparison of the raw image, the PiMAE estimated emitters and the actual emitters, and the comparison of the actual PSF and the PiMAE-estimated PSF. \textbf{b.} NRMSE of the estimated PSF.  \textbf{c.} NRMSE of the estimated emitters.}
            \label{fig:res_lines}
        \end{figure*}

        Dense samples can pose challenges for estimating both the PSF and the emitters. We designed emitters with varying densities as outlined in Appendix \ref{Randomlines}, and employed $\text{LG}_{22}$ as the PSF. As shown in Figure \ref{fig:res_lines}, we observe that as the number of lines in each image ($512 \times 512$) increases, PiMAE's performance in estimating both the PSF and emitters deteriorates. Intuitively, when the number of lines in each image is less than or equal to 50, PiMAE performs well, while performance is poor when the number of lines is greater than 50. This process allows us to assess the influence of sparsity in the emitters on PiMAE.

    \subsection{Computational Resource and Speed}
        In this work, the code is written using the Python library \textit{PyTorch}. PyTorch is a prominent open-source deep learning framework that offers an efficient and user-friendly platform for building and deploying deep learning models. In terms of model training, we utilize three Nvidia Tesla A100 40GB graphics cards in parallel, which is necessary due to ViT's substantial computational and memory requirements. The training time for each task is 11 hours and the inference time for a single $512 \times 512$ image is approximately 4s with the trained model. Compared to supervised models such as DeepSTORM, which takes about 1 hour for training and 0.1s for inference, PiMAE is slower but more powerful.
    
        \hfill

\section{Discussion}
    In this study, we introduce PiMAE, a novel approach for estimating PSF and emitters directly from raw microscopy images. PiMAE addresses several challenges: it allows for direct identification of the PSF from raw data, enabling deep learning model training without the need for real-world or synthetic annotation; it has excellent noise resistance; and it is convenient and widely applicable, requiring only about 5 raw images to resolve the PSF and emitters.
    
    Our method, PiMAE, extracts hidden variables from raw data using physical knowledge. By recognizing PSF as a hidden variable in a linear optical system, the underlying physical principle involves the decomposition of raw data through the convolution of the emitters with the PSF. Hidden variables are ubiquitous in real-world experiments, by integrating masked autoencoder and physical knowledge PiMAE provides a framework to solve hidden variables in physical systems through self-supervised learning.

    However, it should be noted that PiMAE is an emitter localization algorithm, which means that it requires a sufficient degree of sample sparsity to perform effectively. We conducted a evaluation using synthetic data experiments, and while PiMAE performed reasonably well, there is still room for improvement. Therefore, future work could focus on further enhancing the robustness of PiMAE for use in low sparsity scenarios.

\section{Conclusion}
    In conclusion, we have presented PiMAE, a novel solution for directly extracting the PSF and emitters from raw optical microscopy images. By combining the principles of optical microscopy with self-supervised learning, PiMAE demonstrates impressive accuracy and noise robustness in synthetic data experiments, outperforming existing methods such as DeepSTORM and the Richardson-Lucy algorithm. Moreover, our method has been successfully applied to real-world microscopy experiments, resolving wide-field microscopy images with various PSFs. With its ability to learn the hidden mechanisms from raw data, PiMAE has a wide range of potential applications in optical microscopy and scientific studies.

\hfill

\noindent \textbf{Acknowledgements}
    The authors would like to thank Drs. Yu Zheng, Yang Dong, Ce Feng and Shao-Chun zhang for fruitful discussions.

\hfill

\noindent \textbf{Disclosures. } The authors declare no conflicts of interest.

\hfill

\appendix
\section{Network architecture}
\label{nn_arch}
    The principle of microscopic imaging is:
    \begin{equation}
        \text{raw image}=\text{noise}(\text{Emitters} \otimes \text{PSF}) + \text{background,}
    \end{equation}
     where the raw image is the result of convolving the emitters and the PSF with the presence of noise and background. To put this principle into practice, we have developed the PiMAE method, which consists of three modules: emitter inference from raw images, PSF generation, and background separation.
    
    \subsection{Emitter inference}
\begin{figure}[htbp]
            \centering
            \includegraphics[width=\linewidth]{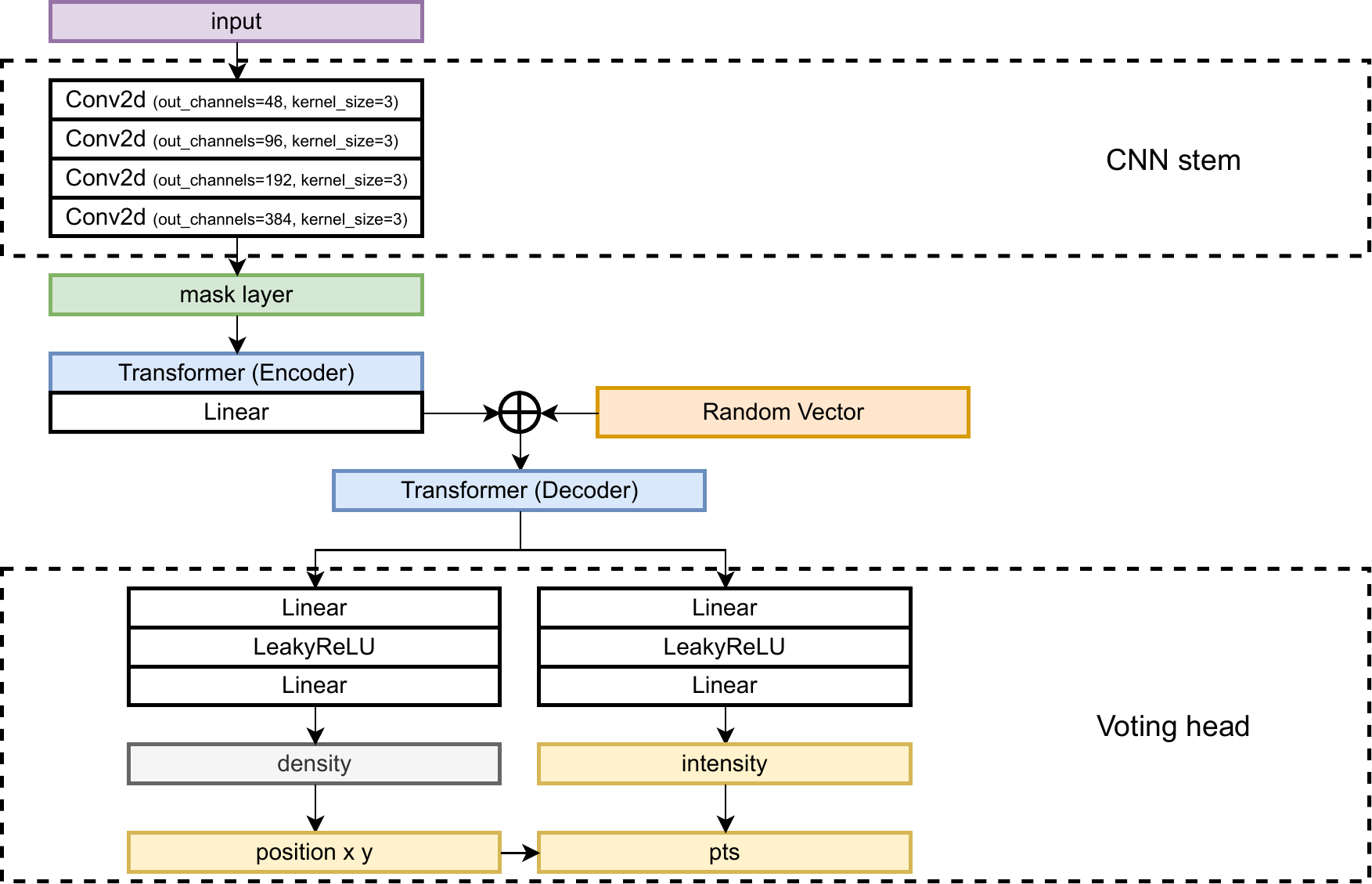}
            \caption{\textbf{Modified masked autoencoder.} CNN stem and voting head are added to the masked autoencoder. The mask layer works during training, masking out 75\% patches.}
            \label{fig:PiMAE-MAE original}
        \end{figure}

        We have improved the original masked autoencoder for use in microscopic imaging by integrating a voting head into its transformer-based decoder. The head predicts the position and intensity of emitters, respectively. Specifically, the decoder produces $9\times9$ feature patches, which serve as the input for the voting head. For emitter position, the voting head employs a two-step process: (1) a multilayer perceptron (MLP) predicts 64 density maps from each feature patch, and (2) the emitter positions are obtained by computing the center of mass of each density map. For emitter intensity, an MLP predicts 64 intensities. The predicted emitter image is generated by placing a Gaussian-type point tensor with $\sigma=1$ scaled by its corresponding intensity at the predicted position, similar to the design in crowd counting methods \cite{zhang2016single}. The mask layer is an essential element in the design of a masked autoencoder. Its main function is to prevent the model from learning trivial solutions and instead encourage it to focus on the relevant features of the input data. This is achieved by randomly blocking out specific parts of the input tensor. To improve the training efficiency, we introduced a CNN stem consisting of four convolutional layers, placed before the mask layer \cite{xiao2021early}. The input image size of $144\times144$ is reduced to $9\times9$ after the CNN stem, with each pixel encoding a 384-dimensional vector. We refer to this model as the point predictor, as shown in Figures \ref{fig:PiMAE-MAE original} and \ref{fig:PiMAE-MAE}.

    \subsection{PSF generation}
        Motivated by the observation that a CNN can function as a well-designed prior and deliver outstanding results in typical inverse problems, as evidenced by Deep Image Prior \cite{ulyanov2018deep}, we constructed the PSF generator as illustrated in Figure \ref{fig:PiMAE-MAE}. The neural network's parameters are adjusted through self-supervised learning to produce the PSF, with a random matrix as the input that remains constant throughout the learning process.
    
    \subsection{Background separation}
\begin{figure}[htbp]
        \centering
        \includegraphics[width=0.8\linewidth]{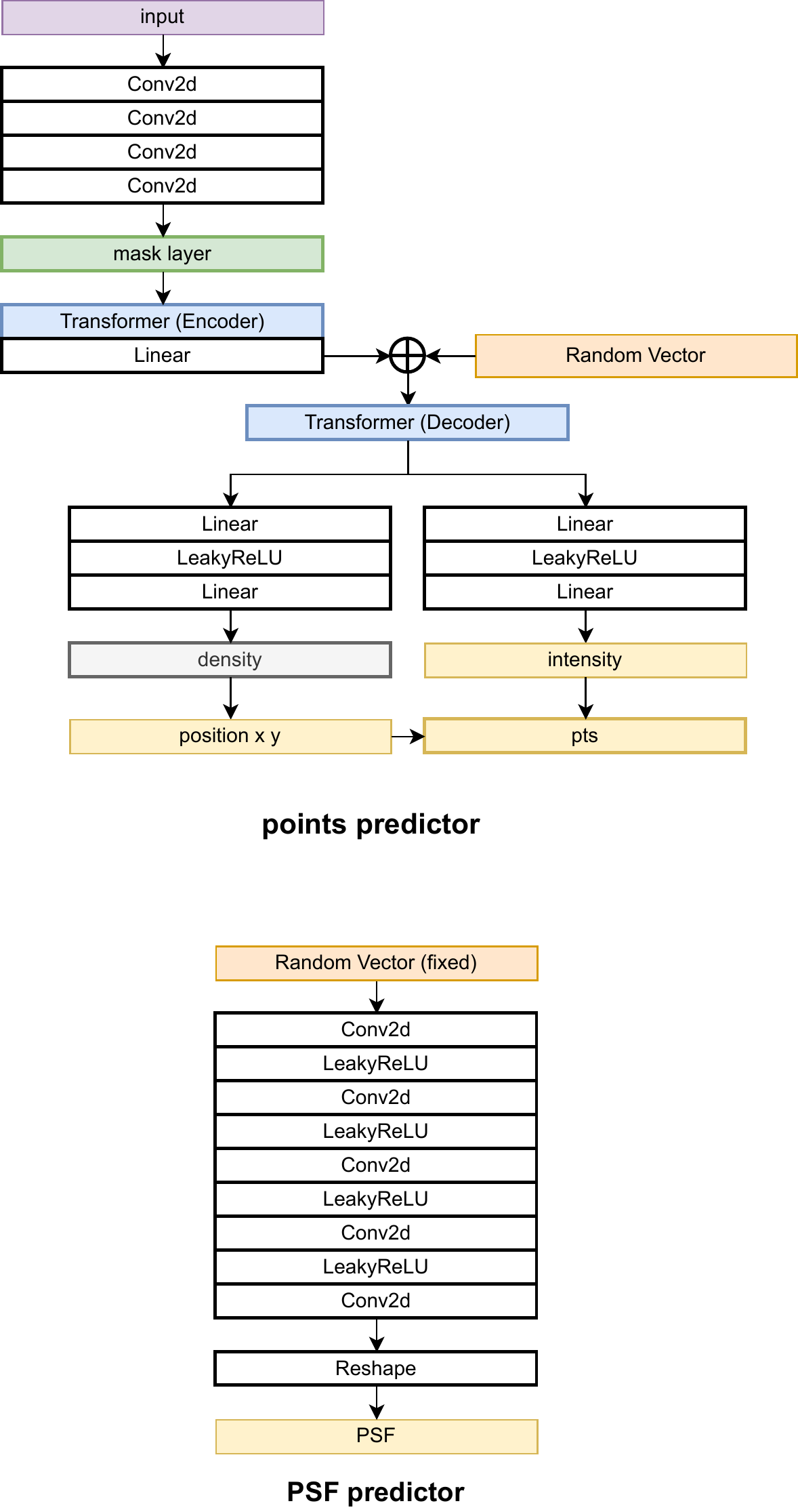}
        \caption{\textbf{Network architecture.} The network consists of two predictors, namely a PSF predictor and a point predictor. The point predictor outputs the location and intensity of the points.}
        \label{fig:PiMAE-MAE}
    \end{figure}

        To isolate the background component from the raw image, we employ a new point predictor (Figure \ref{fig:PiMAE-MAE}). We assume that the background has a low spatial variability and approximate it by drawing the output points from the point predictor following a Gaussian distribution with $\sigma=16$.

\section{Design of loss function}
\label{loss}
    The loss function in our approach is composed of four components, divided into two categories: 
    
    The first category measures the similarity between the reconstructed image and the raw image. It consists of the mean absolute difference (L1) and the multi-scale structural similarity (MS-SSIM), as expressed in Equation \ref{con:msssim}. The combination of these two functions has been demonstrated to perform better than individual functions such as L1 and MSE in image restoration tasks \cite{zhao2016loss}.

    The second category concerns the constraint on the generated PSF. To ensure that the center of mass of the generated PSF is at the center of the PSF tensor, we calculate the center distance loss as follows:
    
    \begin{equation}
    \begin{aligned}
        &\text{Center distance loss} = \\ &\left| \frac{\sum_{i,j} \text{Intensity}{ij} \cdot \overrightarrow{\text{Coordinate}{ij}}}{\sum_{i,j}\text{Intensity}_{ij}} -\overrightarrow{\text{Center position}} \right|,
    \end{aligned}
    \end{equation}
    
    Additionally, to ensure that the generated PSF is spatially continuous, we use the total variation (TV) loss to quantify the smoothness of the image:
    
    \begin{equation}
    \begin{aligned}
    \text{TV loss}& =\\ &\sum_{i,j} (\text{Intensity}_{i, j-1}-\text{Intensity}_{i,j})^2 \\ & +(\text{Intensity}_{i+1, j}-\text{Intensity}_{i,j})^2,
    \end{aligned}
    \end{equation}
    
    Finally, the loss function is defined as:
    
    \begin{equation}
    \begin{aligned}
    &\text{Loss function}= \\ &\alpha_1\text{L1}+\alpha_2\text{MS-SSIM} +\alpha_3\text{Center distance}+\alpha_4\text{TV},
    \end{aligned}
    \end{equation}
    
    where $\alpha_1=0.95$, $\alpha_2=0.05$, $\alpha_3=0.001$ and $\alpha_4=0.001$.

\section{Pretrain with COCO dataset}
\label{Pretrain}
\begin{figure}[htbp]
        \centering
        \includegraphics[width=0.9\linewidth]{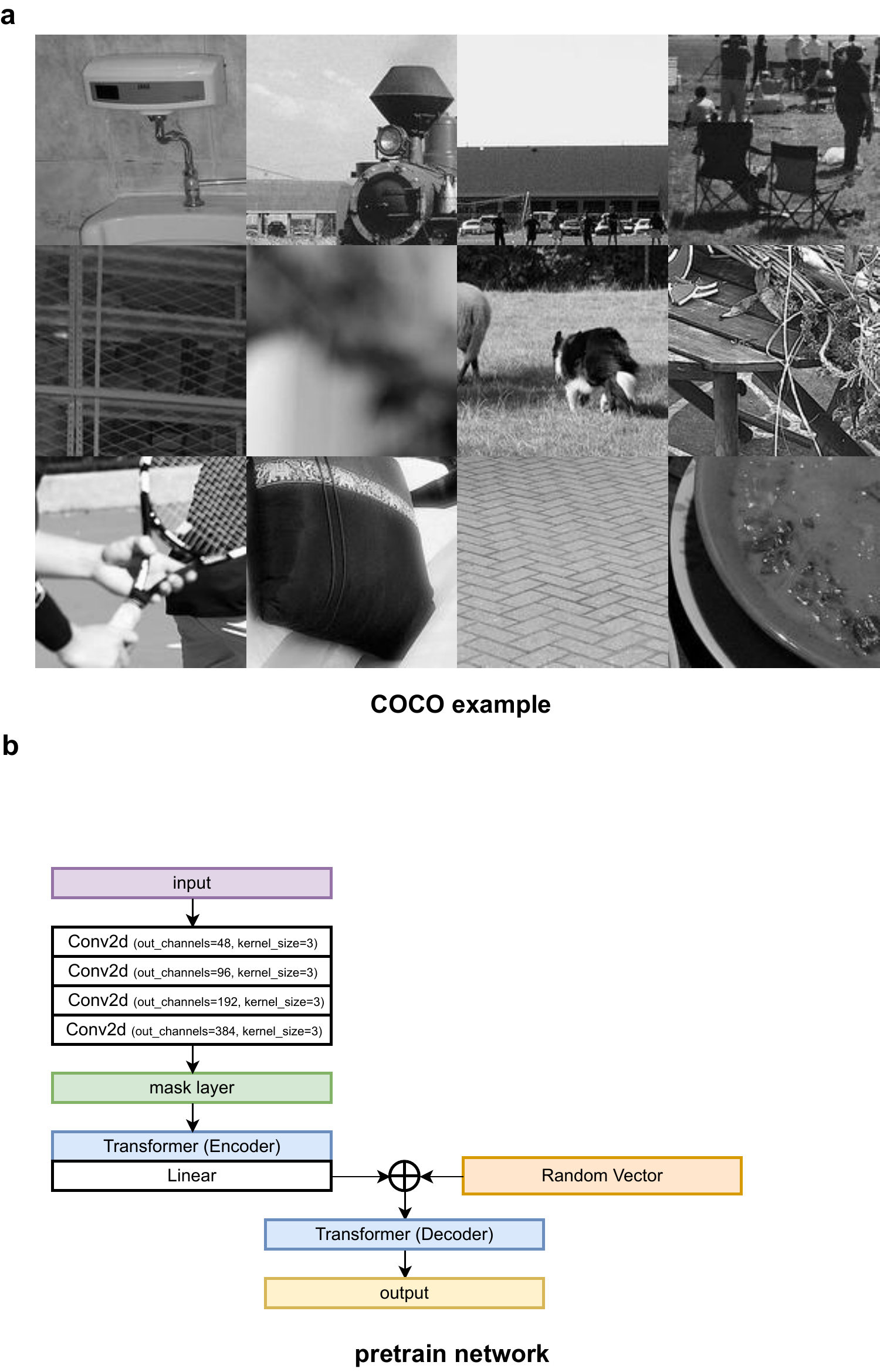}
        \caption{\textbf{Pre-training.} Pre-training with coco dataset. }
        \label{fig:PiMAE-coco}
    \end{figure}

    Recent research has shown that self-supervised pre-training is effective in improving accuracy and robustness in computer vision tasks. In this study, we employed a masked autoencoder (shown in Figure \ref{fig:PiMAE-coco}b) to pretrain the encoder of PiMAE on the COCO dataset \cite{lin2014microsoft} (unlabeled), a large-scale dataset containing 330K RGB images of varying sizes for object detection, segmentation, and captioning tasks.

    To perform pre-training, we randomly cropped $144\times144$ portions from the images and transformed them into grayscale images to form the training set. An example of the cropped images is shown in Figure \ref{fig:PiMAE-coco}a.

\begin{figure}[htbp]
        \centering
        \includegraphics[width=0.7\linewidth]{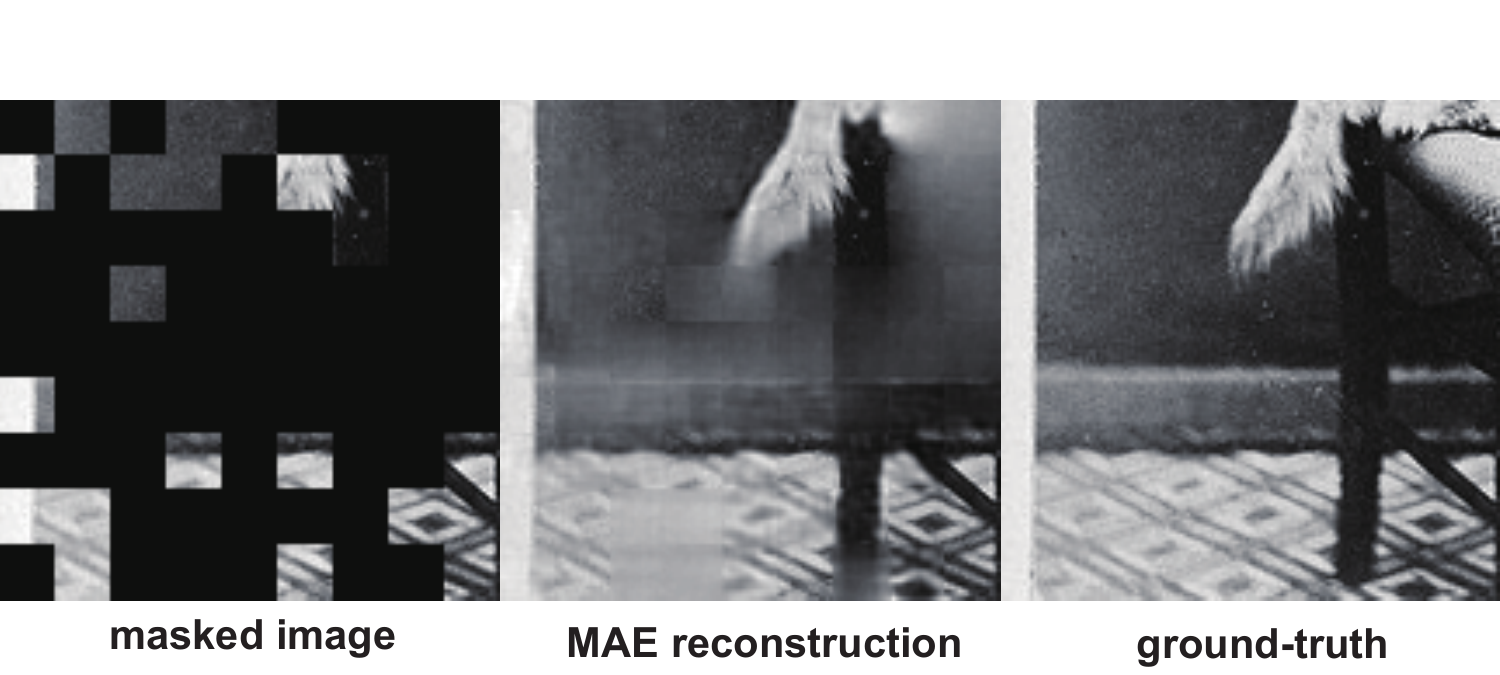}
        \caption{\textbf{Example results on COCO.} We show the masked image, MAE reconstruction, and the ground-truth. The masking ratio here is 0.75.}
        \label{fig:PiMAE-coco-res}
    \end{figure}

\begin{figure}[htbp]
        \centering
        \includegraphics[width=0.7\linewidth]{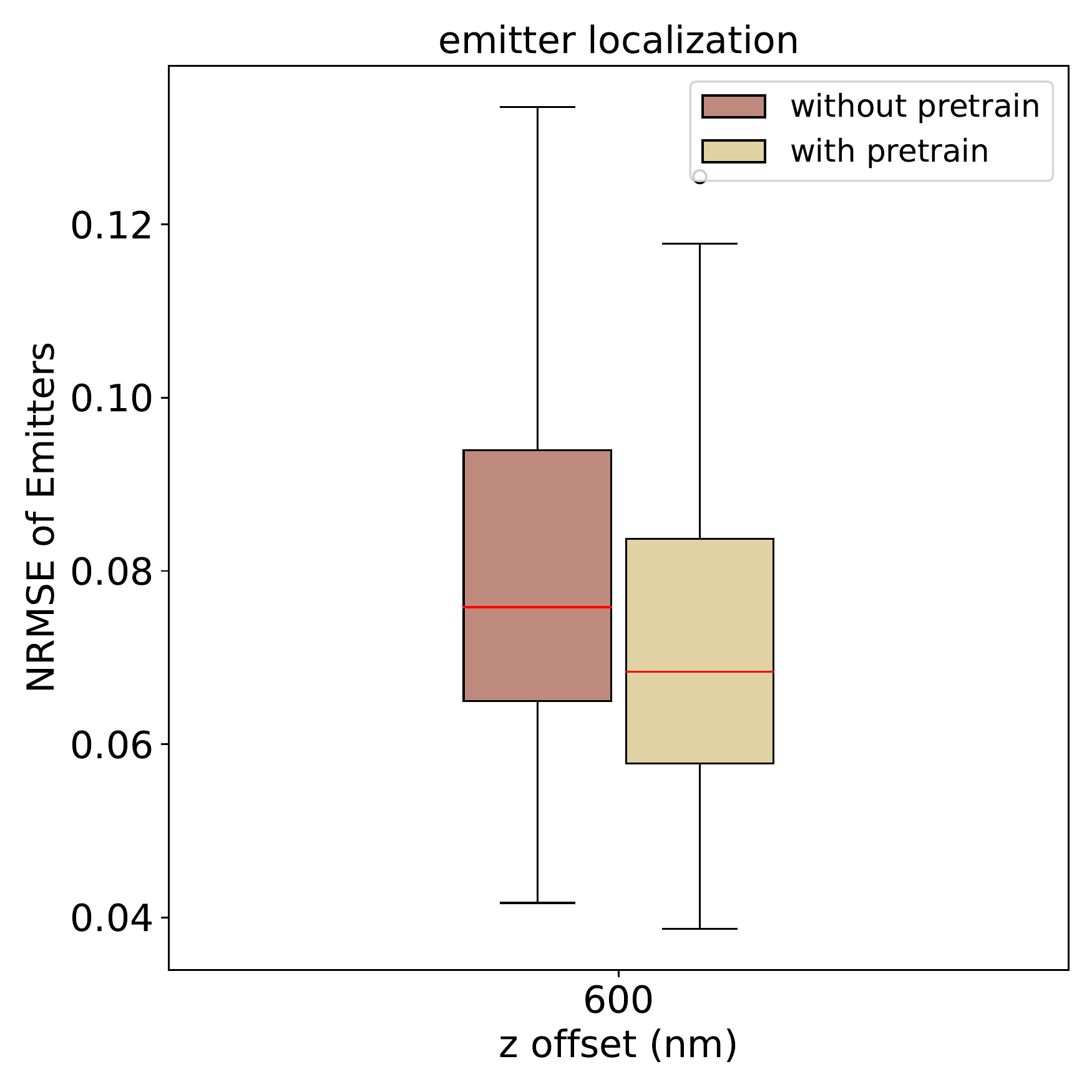}
        \caption{\textbf{Pre-training enhancements.} Comparison of NRMSE metrics for emitter localization of pre-trained and non-pre-trained models. Using 600 nm out-of-focus data as an example, after 500 rounds of training, the learning rate is $3e^{-4}$.}
        \label{fig:PiMAE-pretrain}
    \end{figure}

    We use the mean squared error (MSE) as the loss function during the training process, with a learning rate of $1e^{-4}$ and 500 training epochs. A masking rate of 75\% is implemented, and the RAdam optimizer is used. The results of the MAE reconstruction after pre-training can be seen in Figure \ref{fig:PiMAE-coco-res}. Figure \ref{fig:PiMAE-pretrain} demonstrates that the pre-training process has significantly contributed to the enhancement of the localization of emitters' performance. After completing MAE pre-training, the parameters of the encoder and decoder are saved for the subsequent training of PiMAE.
    
\section{Deconvblind}
    The Deconvblind is one of the most popular methods for blind deconvolution, which iteratively updates the PSF and the estimated image. For each task, we used the training set consisting of 1000 images and applied the Deconvblind function in MATLAB \cite{matlab} to estimate the PSF. These 1000 images were provided to Deconvblind in the form of a stack.
    \subsection{The problem of obtaining a trivial solution in Deconvblind}
    \label{trival}
\begin{figure*}[htbp]
            \centering
            \includegraphics[width=\linewidth]{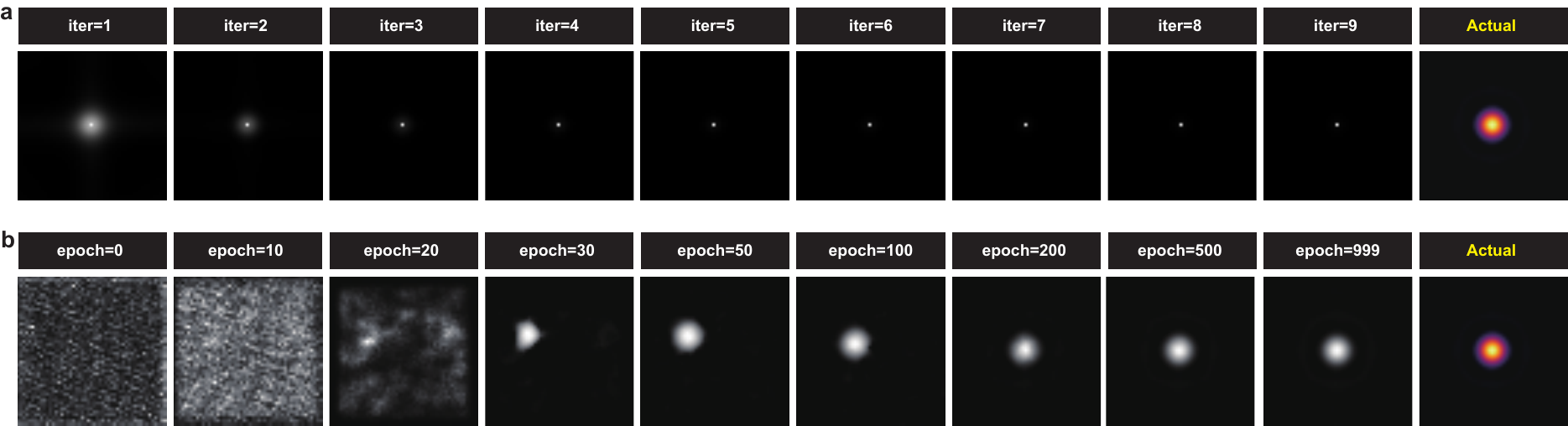}
            \caption{\textbf{Iterative optimization in Deconvblind and PiMAE.} \textbf{a.} The Deconvblind-estimated PSF. \textbf{b.} The PiMAE-estimated PSF.}
            \label{fig:trival}
        \end{figure*}

        Here, we demonstrate that the Deconvblind approach leads to a trivial solution, i.e., a $\delta$-function, for estimating the PSF. We evaluate the performance of Deconvblind and PiMAE on 1000 synthetic images generated from the Sketches dataset, where the PSF is generated from a wide-field microscope in focus. As shown in Figure \ref{fig:trival}a, the PSF estimated by Deconvblind converges to a $\delta$-function, which is a trivial solution and results in the estimated emitter image being equal to the raw image. In contrast, the PiMAE-estimated PSF steadily approaches the actual PSF as the number of training epochs increases.

\section{DeepSTORM}
\begin{figure}[htbp]
        \centering
        \includegraphics[width=\linewidth]{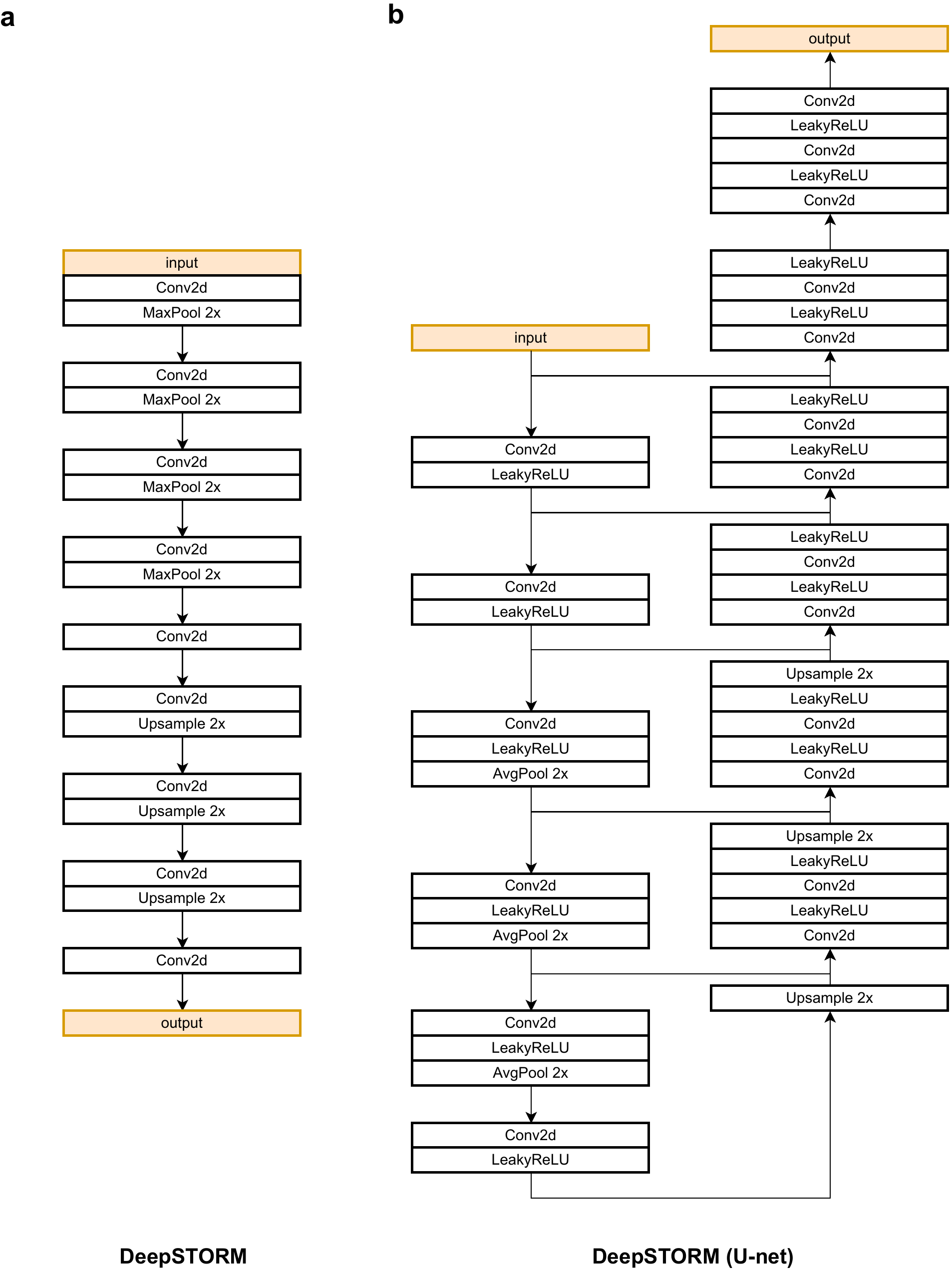}
        \caption{\textbf{Network architecture.} \textbf{a.} Original DeepSTORM architecture. \textbf{b.} Modified DeepSTORM architecture.}
        \label{fig:DeepSTORM-net}
    \end{figure}

    We compare the performance of PiMAE with other deep learning-based methods, such as DeepSTORM, DECODE, and those that train neural networks for predicting emitter locations using supervised learning. As a baseline for comparison, we reproduce the DeepSTORM method. The original DeepSTORM model is a fully convolutional neural network (FCN), which we upgrade to the U-net architecture \cite{ronneberger2015u}, a powerful deep learning architecture that has shown superior performance in various computer vision tasks (see Figure \ref{fig:DeepSTORM-net}). While incorporating this change, we ensure to adhere to the original DeepSTORM model's design and use the sum of MSE and L1 loss as the loss function.

    During the training process, we use 1000 images containing randomly positioned emitters simulated using the ImageJ \cite{collins2007imagej} ThunderSTORM \cite{ovesny2014thunderstorm} plugin. These images are convolved with the PSF of the task, normalized using the mean and averaged standard deviation, and then noise with an intensity of $1e^{-5}$ is added to enhance robustness.

\section{Assesment metrics}
    When evaluating the performance of emitter estimation, we use two metrics: the NRMSE and the Multi-scale Structural Similarity Index (MS-SSIM). NRMSE provides a quantitative measure of the difference between two images, while MS-SSIM is designed to assess the perceived similarity of images, taking into consideration the recognition of emitters by the human eye \cite{wang2003multiscale}.
    
    \textbf{a.} NRMSE defined as:
    \begin{equation}
        \text{NRMSE}=\frac{\sqrt{\sum_{i,j} (\text{Image}_{true} - \text{Image}_{test})^2}}{\text{Max}(\text{Image}_{true})-\text{Min}(\text{Image}_{true})}.
    \end{equation}
    
    \textbf{b.} Multi-scale Structural Similarity (MS-SSIM) defined as:
    \begin{equation}
        \text{MS-SSIM}(\textbf{x},\textbf{y})=[l_m(\textbf{x},\textbf{y})]^{\alpha M} \cdot \prod_{j=1}^{M} [c_j(\textbf{x},\textbf{y})]^{\beta_j}[s_j(\textbf{x},\textbf{y})]^{\gamma_j} \textbf{,}
        \label{con:msssim}
    \end{equation}
    where the exponents $\alpha_m$, $\beta_j$ and $\gamma_j$ are used to adjust the relative importance of different components. Here $\alpha_m=\beta_j=\gamma_j$ and values are 0.0448, 0.2856, 0.3001, 0.2363, 0.1333 for $j=1,2,3,4,5$. The expression of the exponents $l_m$, $c_j$ and $s_j$ are the same as Single-Scale Structural Similarity at each scale $j$,
    \begin{align}
        l(\textbf{x},\textbf{y})&=\frac{2\mu_x \mu_y + C_1}{\mu_x^2+\mu_y^2+C_1}\textbf{,} \\
        c(\textbf{x},\textbf{y})&=\frac{2\sigma_x \sigma_y + C_2}{\sigma_x^2+\sigma_y^2+C_2}\textbf{,} \\
        s(\textbf{x},\textbf{y})&=\frac{\sigma_{xy}+C_3}{\sigma_x \sigma_y+C_3}\textbf{,}
    \end{align}
    where $C_1=(K_1 L)^2$, $C_2=(K_2 L)^2$ and $C_3=C_2/2$,  here $L=255$, $C_1=C_2=0$, $K_1=0.01$ and $K2=0.03$. The sliding window size is 11.

\section{Synthetic Data Generation}
    In this section, we present the construction method of the synthetic data used to evaluate PiMAE, include emitters and PSFs.
    \subsection{Emitters}
        \subsubsection{Sketches}
        \label{Sketches}
\begin{figure}[htbp]
                \centering
                \includegraphics[width=\linewidth]{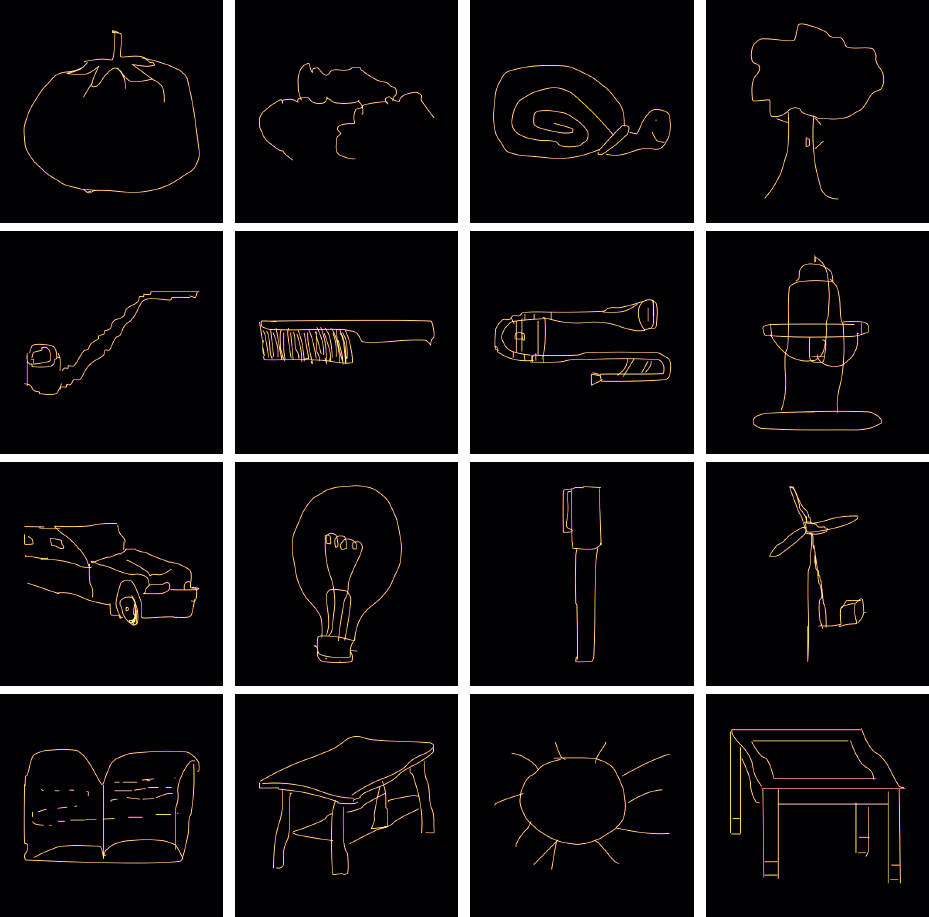}
                \caption{\centering \textbf{Sketches dataset examples.}}
                \label{fig:sketches}
            \end{figure}

            Sketch\cite{eitz2012hdhso} is a large-scale exploration of human sketches containing a rich variety of morphologies. To evaluate the performance of the method, emitters of synthetic data are sampled from the sketches dataset. Figure \ref{fig:sketches} illustrates examples from the Sketch dataset.
        \subsubsection{Random lines}
        \label{Randomlines}
            To evaluate the performance of the model under various levels of sparsity, we implement an algorithm to generate images containing N randomly generated lines:

            \begin{itemize}
                \item[1)] A black image of size $512\times512$ is created.
                \item[2)] A loop is executed N times to randomly draw lines on the image. In each iteration:
                \begin{itemize}
                    \item[a)] The starting and ending points of a line are randomly generated.
                    \item[b)] The intensity of the line is randomly generated.
                    \item[c)] The line is drawn on the image.
                \end{itemize}
                \item[3)] The image is smoothened to remove jaggedness.
            \end{itemize}
            
\begin{figure}[htbp]
                \centering
                \includegraphics[width=\linewidth]{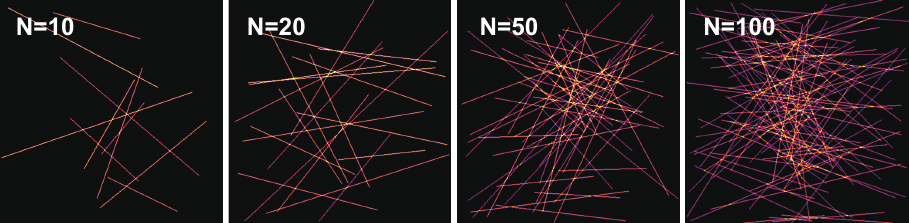}
                \caption{\centering \textbf{randomly generated lines.}}
                \label{fig:lines}
            \end{figure}

            The resulting emitters are shown in the Figure \ref{fig:lines}.
    \subsection{PSFs}
        \label{PSFs}
        \subsubsection{Out-of-focus}
            We simulate the imaging results of a wide-field microscope when the sample is out-of-focus. The near-focus amplitude can be described using the scalar Debye integral\cite{zhang2007gaussian},
            \begin{equation}
                \textbf{h}(x,y,z;\lambda)=C_0\int_0^{\alpha} \sqrt{\text{cos}\theta} \mathcal{J}_0(k\rho\text{sin}\theta) e^{-ikz\text{cos}\theta}\text{sin} \theta d \theta \textbf{,}
            \end{equation}
            where $C_0$ is a complex constant, $\mathcal{J}_0$ is the zero-order Bessel function of the first kind, $\rho=\sqrt{x^2+y^2}$, refractive index is $n$ and numerical aperture $\text{NA}=n\text{sin}\alpha$, wavenumber $k=n (2\pi/\lambda)$. The PSF of the wide-field microscopy is,
            \begin{equation}
                \text{PSF}(x,y,z)=\left| \textbf{h}(x,y,z;\lambda_{em}) \right|^2\textbf{,}
            \end{equation}
            The values of the parameters in this experiment are $C_0=1$, $n=1$, $\lambda_{em}=400$ nm, $\text{NA}=0.7$, and each pixel has a size of 39 nm. $\lambda_{em}$ represents the fluorescence emission wavelength.
    
        \subsubsection{SPADE}
            We simulated four scenarios in the  SPADE, corresponding to PSFs as Hermite Gauss modes $\text{HG}_{\text{22}}$, $\text{HG}_{\text{31}}$ and Laguerre Gauss modes $\text{LG}_{\text{11}}$ $\text{LG}_{\text{22}}$ respectively. Here we set the wavelength to 500 nm, the PSF size to $\text{51 pix} \times \text{51 pix}$ and  $\text{15 mm} \times \text{15 mm}$ range, and rescaled to a 39 nm pixel size. The definitions for the amplitude of the Hermite Gauss modes and Laguerre Gauss modes are\cite{beijersbergen1993astigmatic},
            
            \begin{equation}
                \begin{aligned}
                    u_{nm}^{HG} (x,y,z) &= C_{nm}^{HG}(1/w)\text{exp}[- ik(x^{2}+y^{2}/2R)] \\
                    & \times\text{exp}[-(x^{2}+y^{2}/w^{2})]\text{exp}[-i(-n+m+1)\psi] \\
                    & \times H_{n}(x\sqrt 2/w)H_{m}(y\sqrt 2/w) \textbf{,}
                \end{aligned}
            \end{equation}
            
            \begin{equation}
                \begin{aligned}
                    u^{LG}_{nm}(r,\phi,z) &= C ^{LG}_{nm}(1/w)\text{exp}(-ikr^{2}/2R)\text{exp }(-r^{2}/w^{2}) \\
                    & \times \text{exp}[-i(n + m + 1)\psi ]\text{exp}\left[-i(n-m)\phi\right] \\
                    & \times (-1)^{\text{min}(n,m)}(r\sqrt2/w)^{\left| r-m \right|} \\
                    & \times L^{\left| n-m \right|}_{\text{min}(n,m)}(2r^2/w^2) \textbf{,}
                \end{aligned}
            \end{equation}
            with $R(z)=(z^{2}_{R}+z^{2})/z_{R}$, $\frac{1}{2}kw^{2}(z)=(z^{2}_{R}+z^{2})/z_{R}$, and $\psi(z)=\text{arctan}(z/z_{R})$. $H_n(x)$ is the Hermite polynomial of order $n$, $L_p^l(x)$ is the generalized Laguerre polynomial, $k=\frac{2\pi}{\lambda}$ is the wave number, $z_R$ is the Rayleigh range of the mode. Here we set $w_0=\text{2 mm}$, wavelength $\lambda=\text{500 nm}$ and $z=0$.

\section{Data normalization}
    We use the Max-Min normalization method to process each image as follows:
    \begin{equation}
        x_{\text{norm}}=\frac{x - x_{\text{min}}}{x_{\text{max}} - x_{\text{min}}}
    \end{equation}
    where $x_{norm}$ is the normalized image, $x$ is the raw image, $x_{min}$ is the minimum value in the image, and $x_{max}$ is the maximum value in the image.

\section{Evaluation results of adding speckle noise to synthetic data}
\begin{figure*}[htbp]
        \centering
        \includegraphics[width=0.80\linewidth]{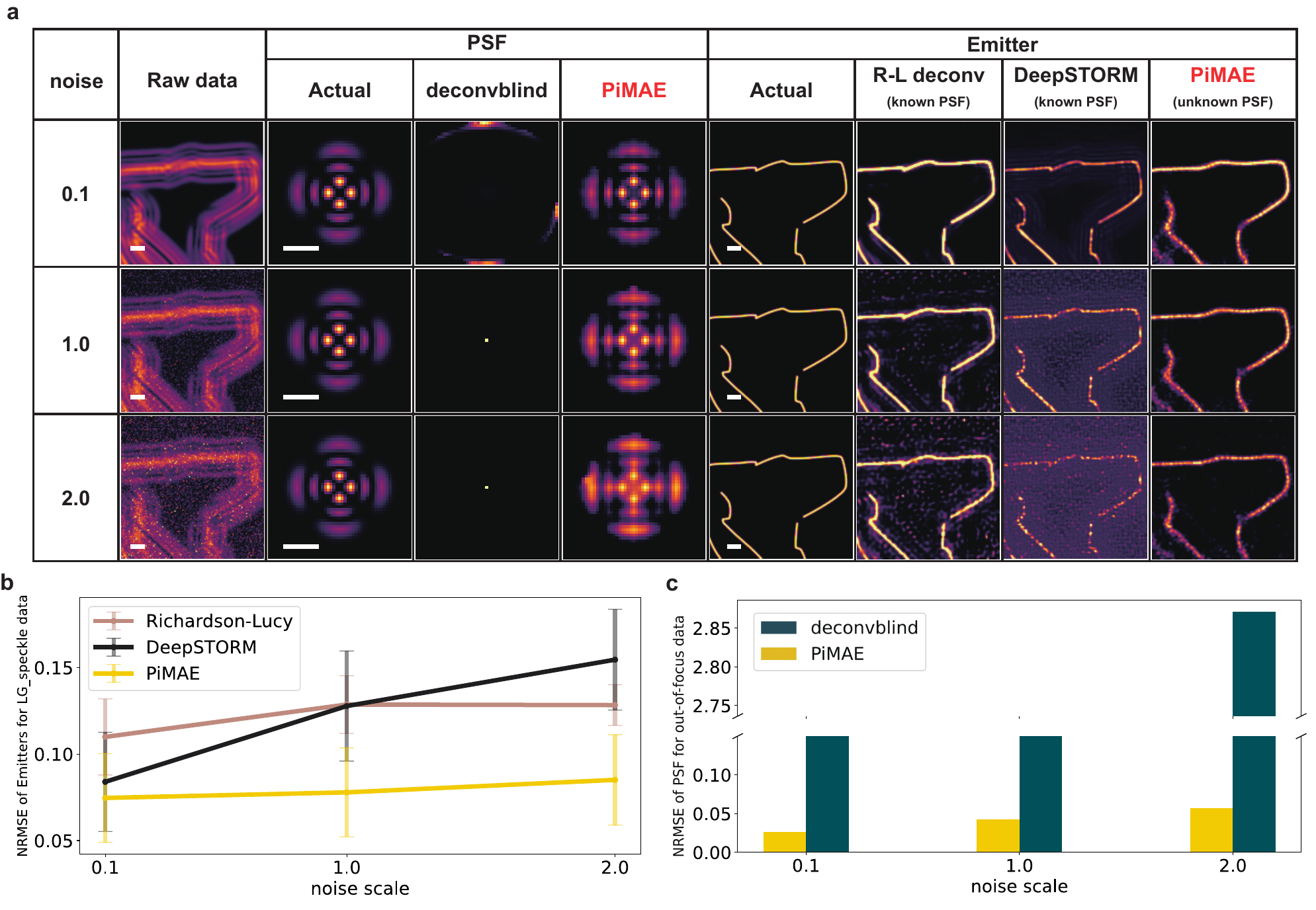}
        \caption{\textbf{Evaluation of speckle noise robustness.} The scale bar is 0.5 \textmu m. \textbf{a.} The estimated PSF and emitters results from synthetic data with speckle noise. \textbf{b.} NRMSE of the results of estimated PSF from synthetic data with speckle noise. \textbf{c.} NRMSE of the results of estimated emitters synthetic data with speckle noise.}
        \label{fig:speckle_noise}
    \end{figure*}

    Speckle noise is a type of granular noise texture that can degrade image quality in coherent imaging systems such as medical ultrasound, optical coherence tomography, as well as radar and synthetic aperture radar (SAR) systems. It is a multiplicative noise that is proportional to the image intensity. The probability density function of speckle noise can be described by an exponential distribution:
    \begin{equation}
        p(z) = \frac{1}{\sigma^2} \exp \left( -\frac{z}{\sigma^2} \right)
    \end{equation}
    Here, $z$ represents the intensity, and $\sigma^2$ represents the speckle noise variance. To evaluate the impact of speckle noise on estimating PSF and emitters, we use $\text{LG}_{22}$ as the PSF and Sketches as the emitters. We construct three sets of data with noise variances of 0.1, 1, and 2, respectively, each containing 1000 training images and 100 test images. We use the NRMSE metric to evaluate the results, as shown in Figure \ref{fig:speckle_noise}.

\section{The results using MS-SSIM as the metric}
    \subsection{Results of out-of-focus synthetic data}
\begin{figure}[htbp]
            \centering
            \includegraphics[width=0.75\linewidth]{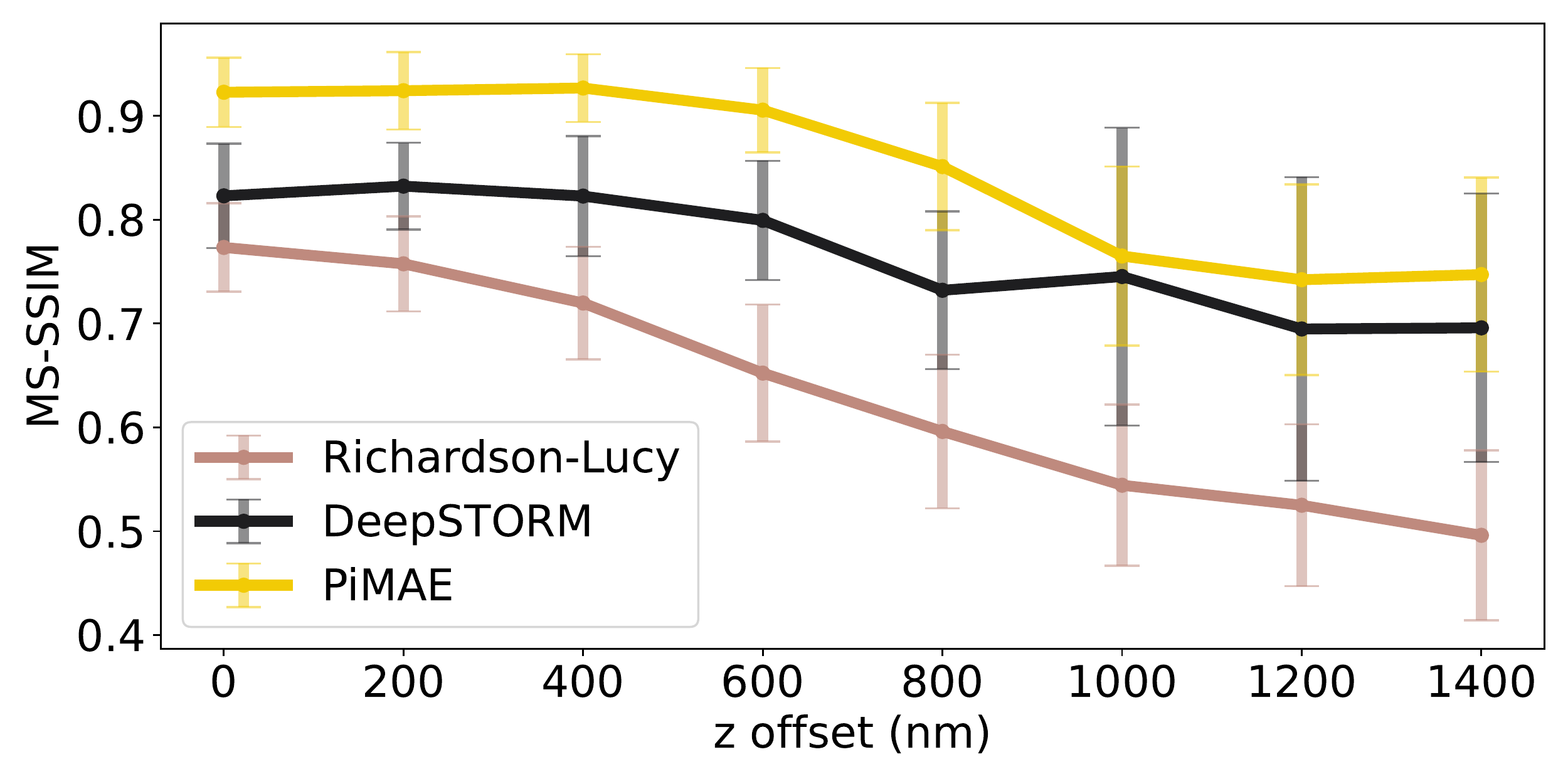}
            \caption{MS-SSIM of the results of estimating emitters from out-of-focus synthetic data.}
            \label{fig:sketches_z_off_msssim}
        \end{figure}

        In this section, we present the results of synthetic data with varying out-of-focus distances, assessed using the MS-SSIM metric. Gaussian noise with a standard deviation of $\text{noise}_{\text{std}}/\text{raw}_{\text{mean}}=0.5$ was added to each synthetic data set. The results are displayed in Figure \ref{fig:sketches_z_off_msssim}.
    \subsection{Results of SPADE synthetic data}
\begin{figure}[htbp]
            \centering
            \includegraphics[width=0.75\linewidth]{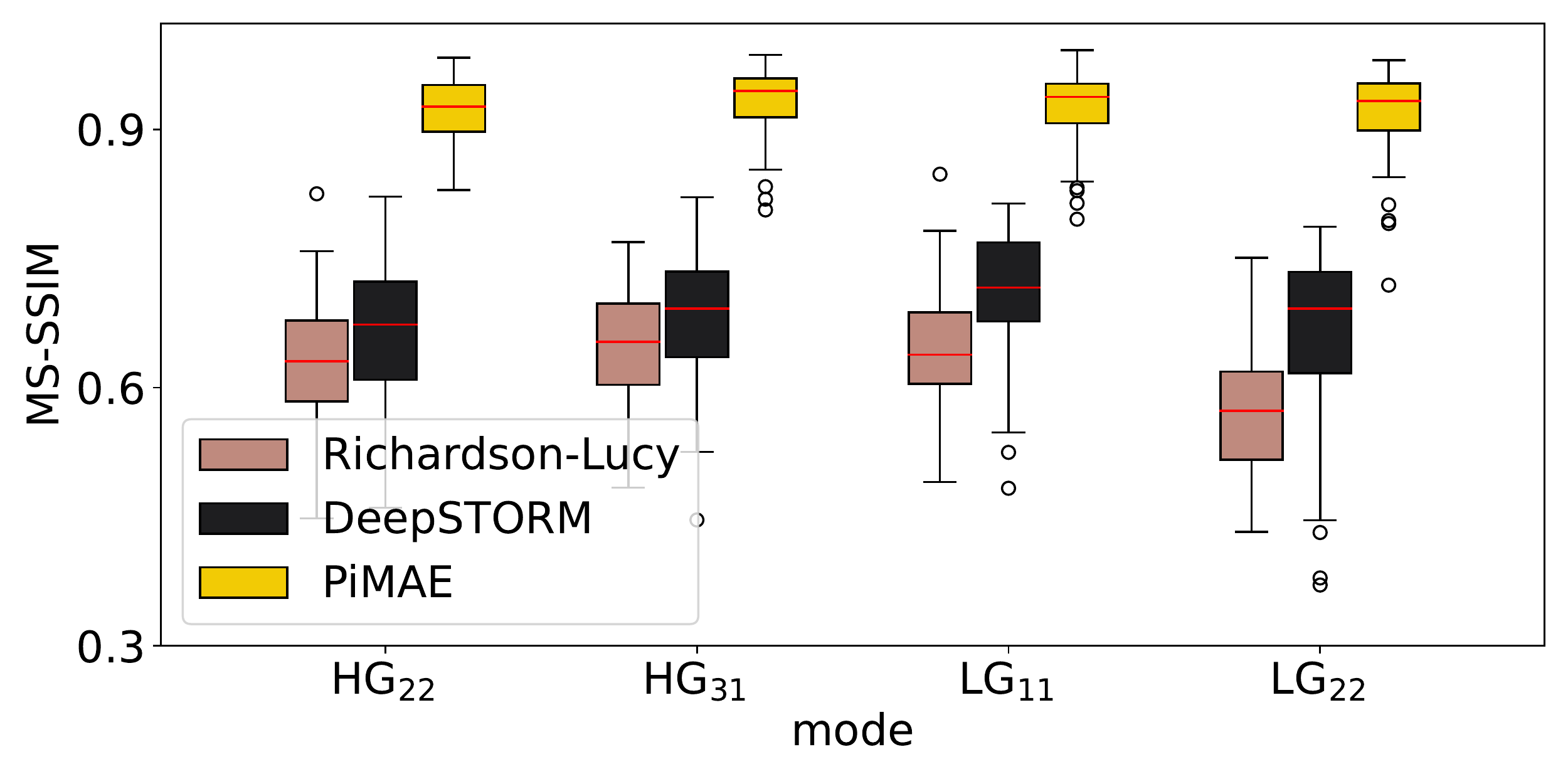}
            \caption{MS-SSIM of the results of estimating emitters from the "SPADE" Sketches dataset.}
            \label{fig:sketches_SPADE_msssim}
        \end{figure}

        We present the results for synthetic data evaluated using the SSIM metric for Hermite-Gaussian (HG) and Laguerre-Gaussian (LG) modes. Gaussian noise with $\text{noise}_{\text{std}}/\text{raw}_{\text{mean}}=0.5 $ is added to each synthetic data set. The results are displayed in Figure \ref{fig:sketches_SPADE_msssim}.
    \subsection{Results of noise robustness}
\begin{figure}[htbp]
            \centering
            \includegraphics[width=0.75\linewidth]{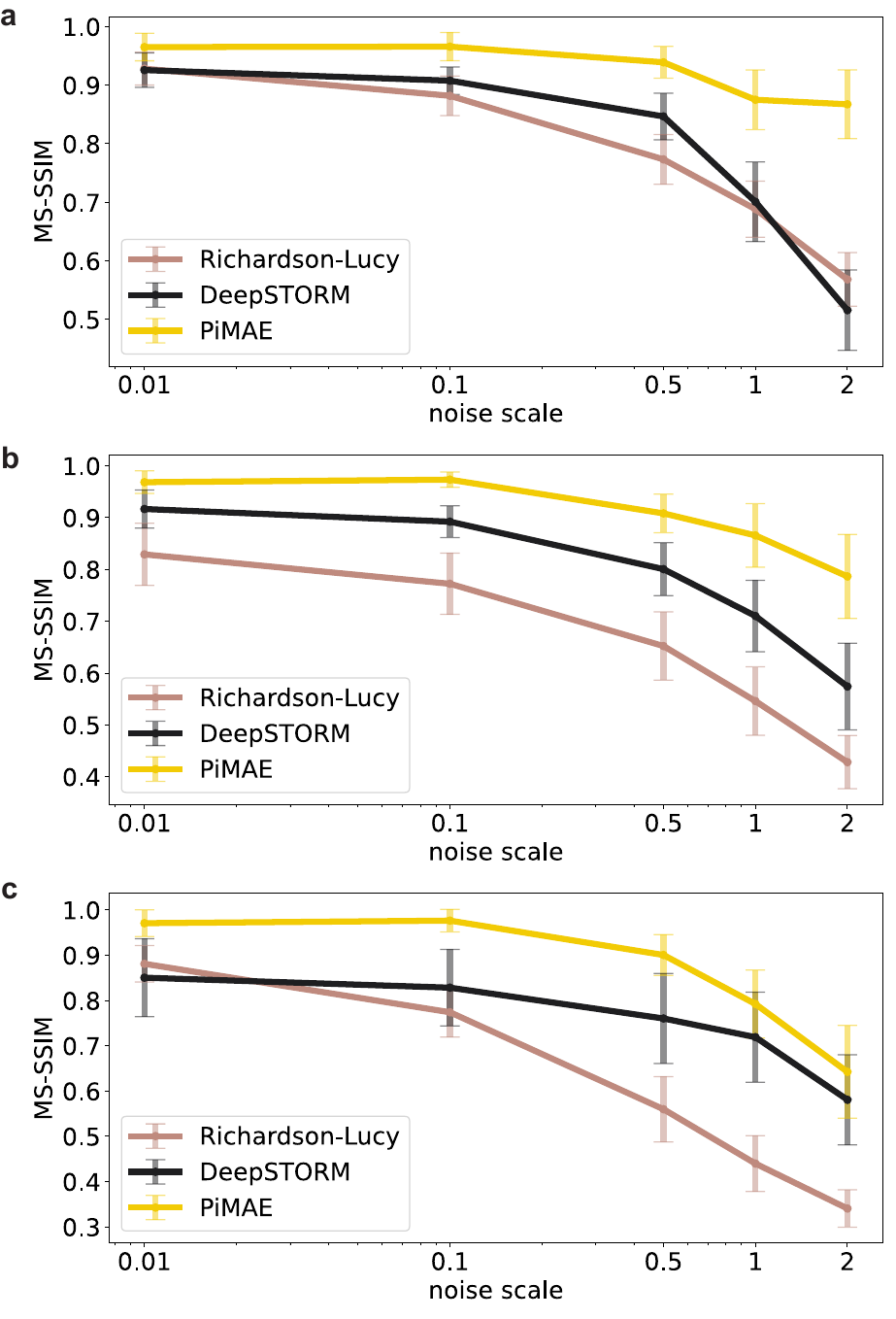}
            \caption{\textbf{Noise robustness.} \textbf{a.} MS-SSIM of the results of estimated emitters from the in-focus Sketches dataset. \textbf{b.} MS-SSIM of the results of estimated emitters from the 600 nm out-of-focus Sketches dataset. \textbf{c.} MS-SSIM of the results of estimated emitters from the $\text{LG}_\text{22}$ mode Sketches dataset.}
            \label{fig:sketches_noise_msssim}
        \end{figure}

        We present the results of synthetic data with different levels of noise measured using SSIM as the metric. For each synthetic dataset, Gaussian noise was added with levels of 0.01, 0.1, 0.5, 1, and 2, respectively.The results are depicted in Figure \ref{fig:sketches_noise_msssim}.
    
\section{Results of real-world experiments}
    We assess the efficacy of PiMAE in two real-world experiments. Firstly, we utilize the imaging results of endoplasmic reticulum (ER) structures obtained from both wide-field microscopy and structural illumination microscopy (SIM) from the BioSR dataset \cite{qiao2021evaluation}. Secondly, we construct a custom-built wide-field microscope to image nitrogen vacancy color centers in diamond. The ability of PiMAE to handle non-Gaussian PSFs is evaluated in both out-of-focus and aberrations scenarios.
    
    \subsection{Results of endoplasmic reticulum}
\begin{figure*}[htbp]
            \centering
            \includegraphics[width=0.9\linewidth]{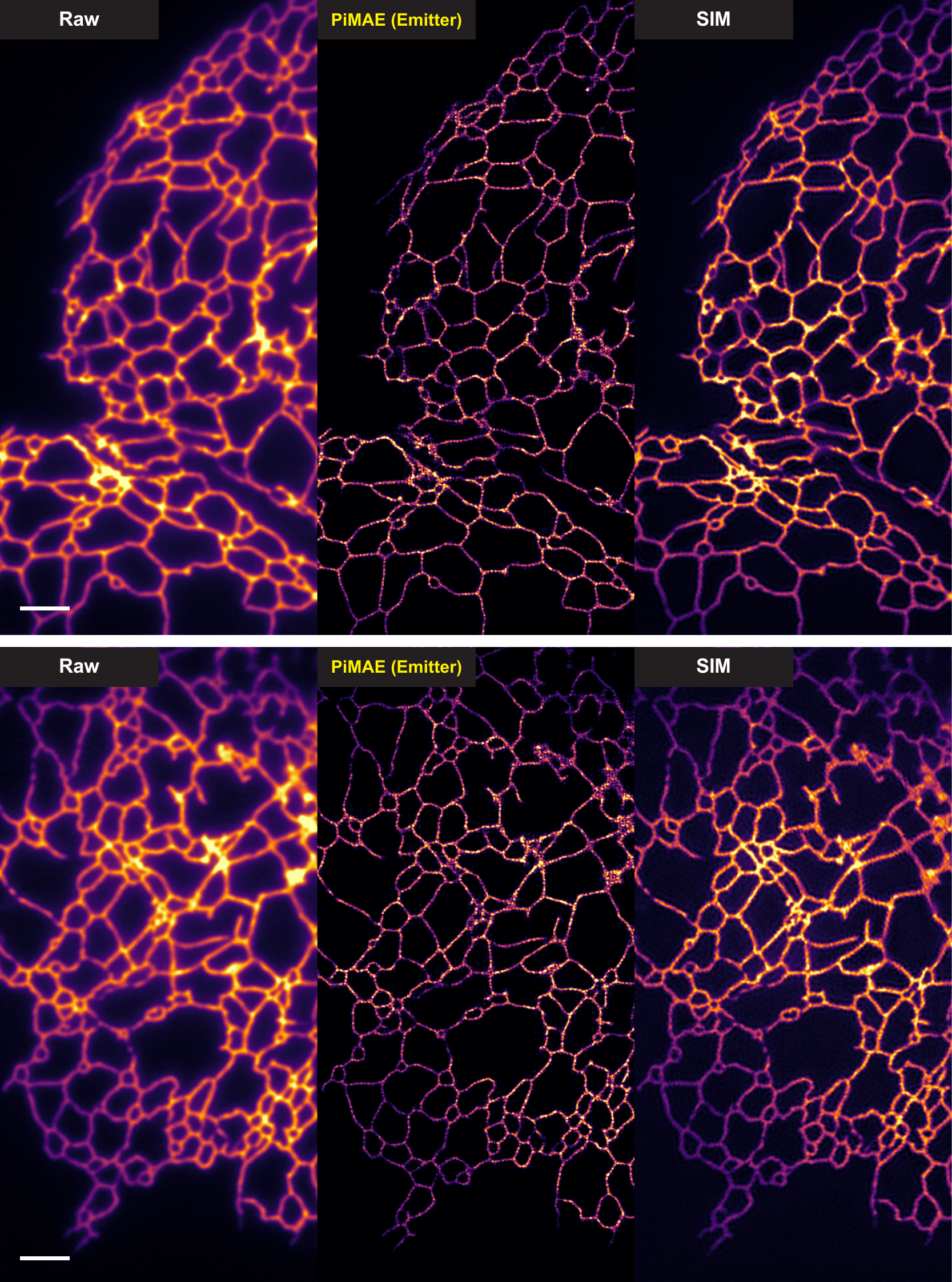}
            \caption{\textbf{Comparison of endoplasmic reticulum imaging results}. Here we show some comparative results of wide-field microscopy, SIM and PiMAE-resolved wide-field microscopy. The length of the scale bar is 2.50 \textmu m. Data from BioSR dataset.\cite{qiao2021evaluation}}
            \label{fig:biosr_er_exp}
        \end{figure*}

\begin{figure*}[htbp]
            \centering
            \includegraphics[width=0.9\linewidth]{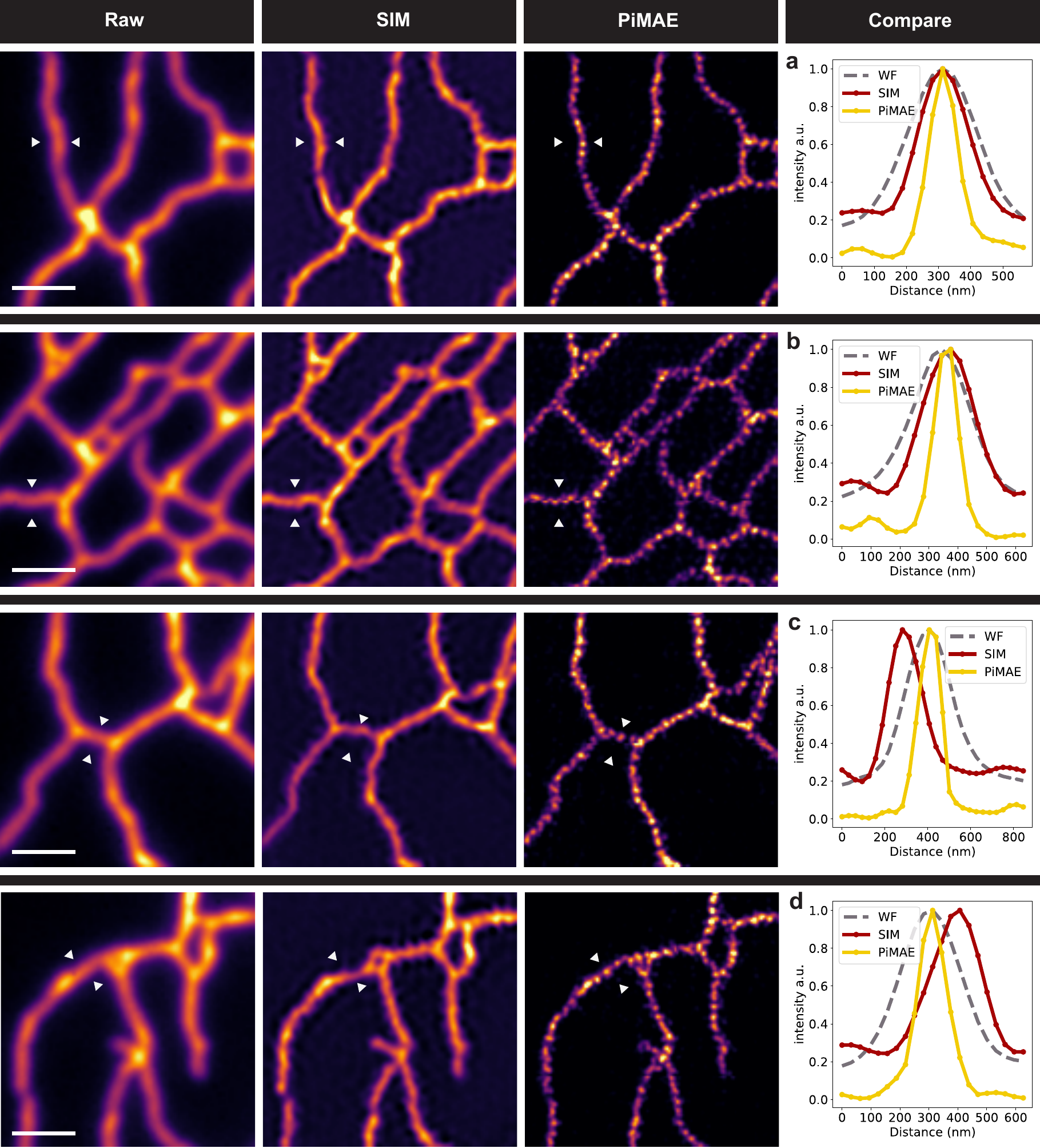}
            \caption{\textbf{Artifacts in super-resolution images reconstructed using SIM.} Reconstruction artifacts are a common issue in SIM-reconstructed images, as evidenced in Figures c-d, due to factors such as non-uniform fringe patterns or phase errors in the reconstruction process. In comparison, the PiMAE-estimated emitters do not exhibit these artifact problems. The length of the scale bar is 1.00 \textmu m.}
            \label{fig:biosr_er_artifacts}
        \end{figure*}

        Figure \ref{fig:biosr_er_exp} shows the results of wide-field microscopy, SIM and PiMAE-resolved wide-field microscopy of ER and Figure \ref{fig:biosr_er_artifacts}  demonstrates that PiMAE is capable of avoiding the artifact phenomenon seen in SIM.
        
    \subsection{Results of NV center imaging}
\begin{figure*}[htbp]
            \centering
            \includegraphics[width=0.9\linewidth]{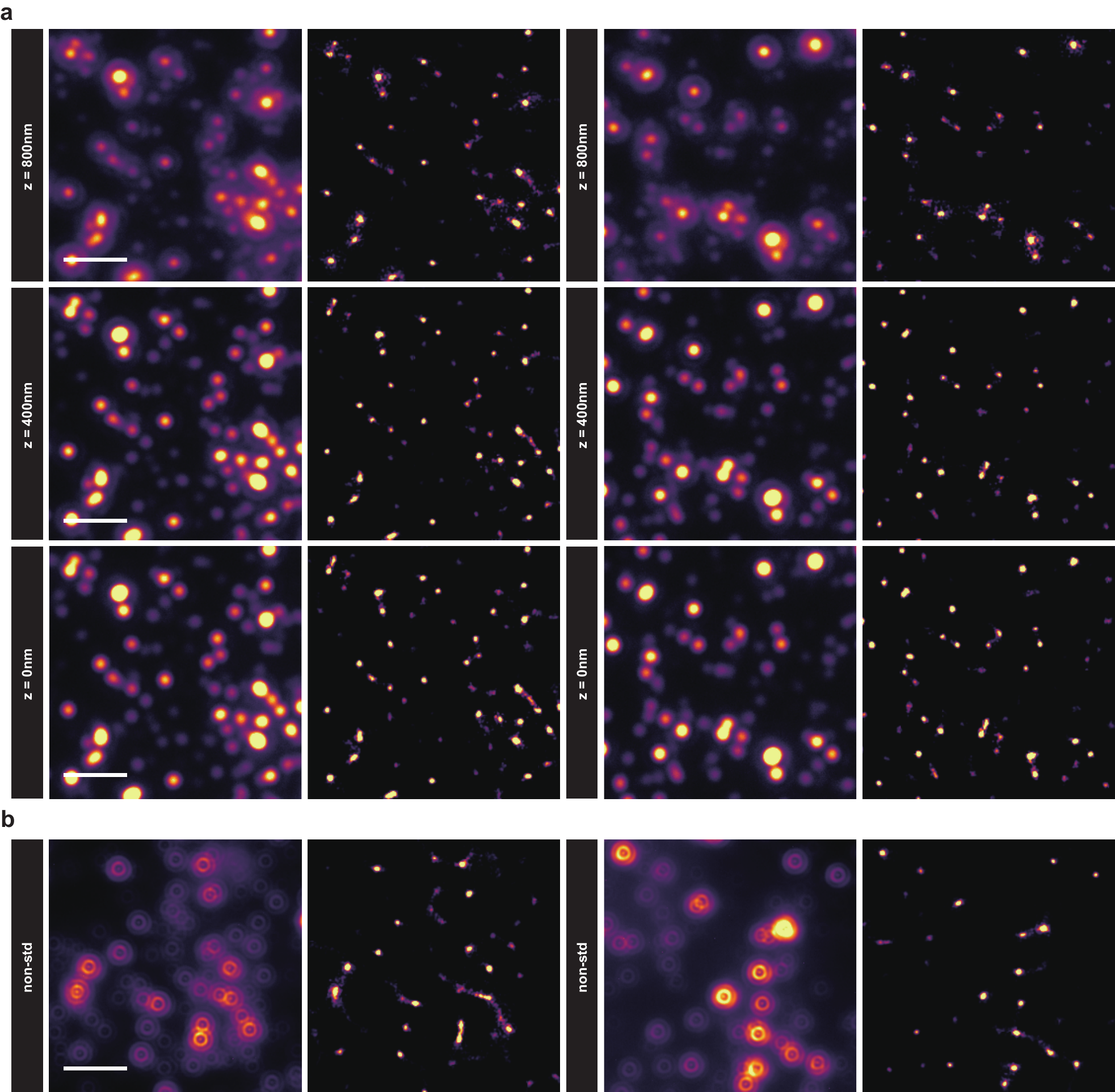}
            \caption{\textbf{Wide-field microscopy imaging of NV color center.} \textbf{a.} The comparison of wide-field microscopy results and PiMAE estimated emitter results at different out-of-focus distances, with invariant field of view from top to bottom, and different field of view on the left and right, respectively. The length of the scale bar is 2.50 \textmu m. \textbf{b.} The wide-field microscopy results and PiMAE estimated emitters of non-standard PSF when the objective is mismatched to the coverslip. The length of the scale bar is 6.40 \textmu m.}
            \label{fig:nv_wf}
        \end{figure*}

        The results of out-of-focus and aberrated wide-field microscopy imaging of nitrogen vacancy (NV) color centers, as well as PiMAE-resolved results, are shown in Figure \ref{fig:nv_wf}. The aberrations were generated as follows: an objective lens with a phase aberration correction ring was first used to image a 50 nm nanodiamond on the opposite side of a coverslip with a thickness of 0.11-0.23 mm and a refractive index of 1.5. The correction ring of the objective lens was then rotated to match a coverslip with a thickness of 0.1 mm and a refractive index of 1.5, and this lens was used to observe nanodiamonds spin-coated on the opposite side of sapphire with a thickness of 0.15 mm and a refractive index of 1.72, thus artificially creating an aberration and resulting in a doughnut-shaped PSF. (Note: Olympus UPLXAPO40X objective lens was used)

\section{Train set size}
\label{training_set_size}
\begin{figure*}[htbp]
        \centering
        \includegraphics[width=0.80\linewidth]{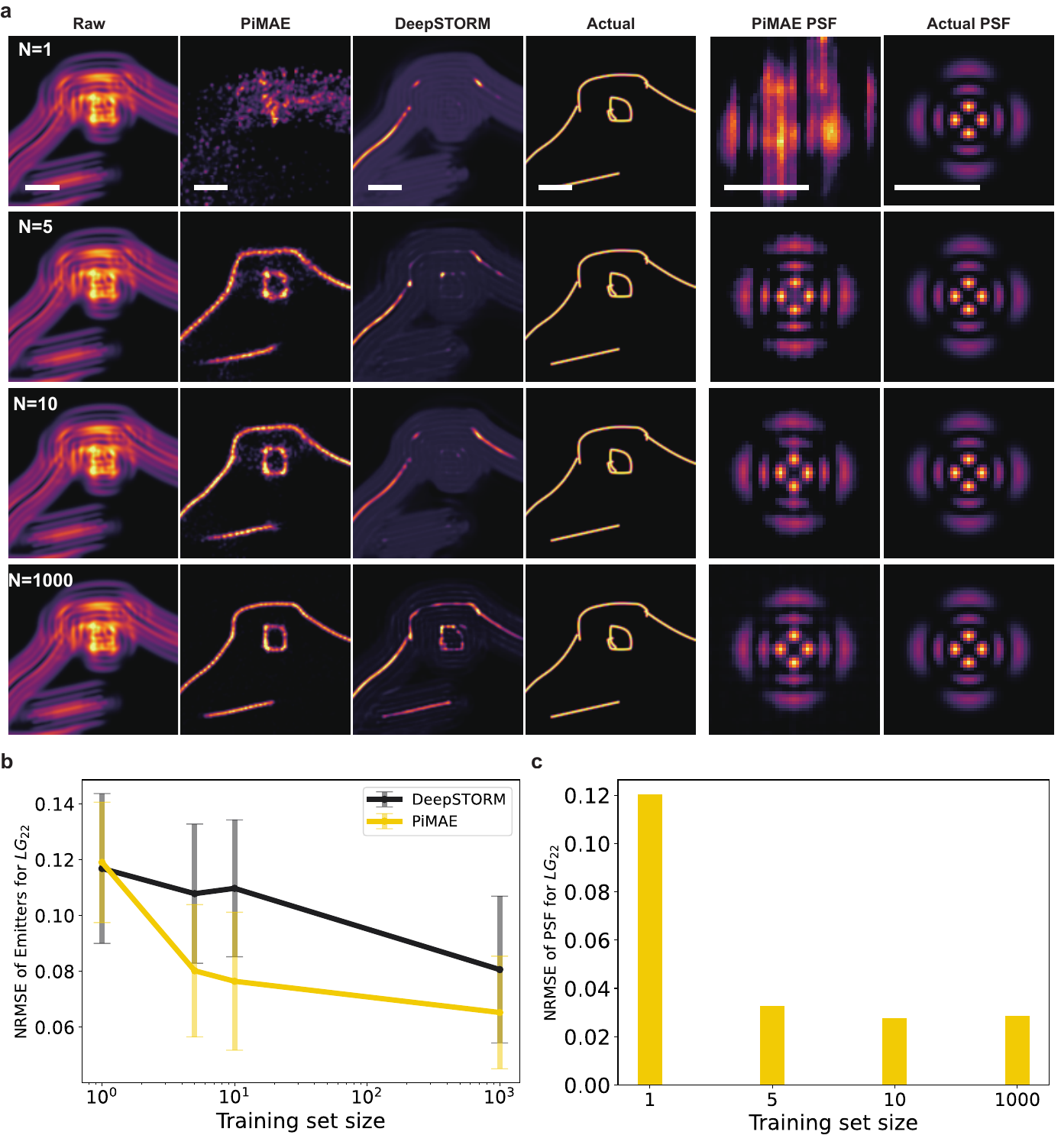}
        \caption{\textbf{Evaluating the Effect of Training Set Size.} The scale bar is 1.0 \textmu m. \textbf{a.} Results of estimated PSF and emitters when the size of the training set is 1, 5, 10, and 1000, and the size of the test set is 100. \textbf{b.} NRMSE of the results of estimated emitters from synthetic data with different dataset sizes. \textbf{c.} NRMSE of PiMAE-estimated PSF results from synthetic data with different dataset sizes.}
        \label{fig:datasize}
    \end{figure*}

    We use $\text{LG}_{22}$ as the PSF and a fixed test set size of 100 images with a shape of $512\times512$. The training set sizes for both PiMAE and DeepSTORM are 1, 5, 10, and 1000 images, respectively. As shown in Figure \ref{fig:datasize}, PiMAE performs well even with a training set size as small as 5 images, whereas the performance of DeepSTORM decreases significantly.

\section{Summary of results}
    In this section, we summarize the results of all the synthetic tasks in the table \ref{tab:performance},
    \begin{table*}[htbp]
        \centering
        \begin{tabular}{c c c c | c c c | c c}
            \hline
            \multicolumn{9}{c}{Synthetic data}\\

            \hline
            \multicolumn{4}{c|}{Task info} & \multicolumn{3}{c|}{NRMSE for Emitters} & \multicolumn{2}{c}{NRMSE for PSF} \\
            \hline
            Task & PSF & Emitters & Noise & PiMAE & DeepSTORM & RL & PiMAE & DB \\
            \hline
                1    & 1400nm & Sketches & 5.00e-01 & 0.09 & 0.111 & 0.257 & 0.07 & 0.195 \\
                2    & 1200nm & Sketches & 5.00e-01 & 0.09 & 0.106 & 0.238 & 0.075 & 0.144 \\
                3    & 1000nm & Sketches & 5.00e-01 & 0.093 & 0.11 & 0.232 & 0.083 & 0.098 \\
                4    & 800nm  & Sketches & 5.00e-01 & 0.08 & 0.103 & 0.201 & 0.029 & 0.062 \\
                5    & 600nm  & Sketches & 5.00e-01 & 0.073 & 0.092 & 0.163 & 0.018 & 0.059 \\
                6    & 400nm  & Sketches & 5.00e-01 & 0.074 & 0.081 & 0.14 & 0.018 & 0.051 \\
                7    & 200nm  & Sketches & 5.00e-01 & 0.072 & 0.078 & 0.13 & 0.023 & 0.048 \\
                8    & 0nm    & Sketches & 5.00e-01 & 0.071 & 0.084 & 0.124 & 0.022 & 0.045 \\
                9    & 0nm    & Sketches & 2.00e+00 & 0.089 & 0.139 & 0.198 & 0.045 & 0.078 \\
                10   & 0nm    & Sketches & 1.00e+00 & 0.085 & 0.105 & 0.156 & 0.031 & 0.064 \\
                11   & 0nm    & Sketches & 5.00e-01 & 0.071 & 0.079 & 0.124 & 0.022 & 0.045 \\
                12   & 0nm    & Sketches & 1.00e-01 & 0.068 & 0.066 & 0.091 & 0.021 & 0.042 \\
                13   & 0nm    & Sketches & 1.00e-02 & 0.068 & 0.065 & 0.082 & 0.021 & 0.165 \\
                14   & 600nm  & Sketches & 2.00e+00 & 0.095 & 0.144 & 0.231 & 0.019 & 0.076 \\
                15   & 600nm  & Sketches & 1.00e+00 & 0.091 & 0.111 & 0.185 & 0.016 & 0.07 \\
                16   & 600nm  & Sketches & 5.00e-01 & 0.073 & 0.092 & 0.163 & 0.018 & 0.058 \\
                17   & 600nm  & Sketches & 1.00e-01 & 0.066 & 0.073 & 0.142 & 0.023 & 0.03 \\
                18   & 600nm  & Sketches & 1.00e-02 & 0.068 & 0.07 & 0.135 & 0.023 & 0.937 \\
                19   & HG/2\_2 & Sketches & 5.00e-01 & 0.075 & 0.098 & 0.151 & 0.028 & 0.156 \\
                20   & HG/3\_1 & Sketches & 5.00e-01 & 0.072 & 0.097 & 0.147 & 0.029 & 0.161 \\
                21   & LG/1\_1 & Sketches & 5.00e-01 & 0.072 & 0.098 & 0.154 & 0.016 & 0.088 \\
                22 & LG/2\_2 & Sketches & 5.00e-01 & 0.073 & 0.094 & 0.179 & 0.042 & 0.042 \\
                23   & LG/2\_2 & Sketches & 2.00e+00 & 0.1 & 0.128 & 0.307 & 0.069 & 0.105 \\
                24   & LG/2\_2 & Sketches & 1.00e+00 & 0.078 & 0.104 & 0.235 & 0.048 & 0.1 \\
                25   & LG/2\_2 & Sketches & 5.00e-01 & 0.063 & 0.094 & 0.179 & 0.029 & 0.098 \\
                26   & LG/2\_2 & Sketches & 1.00e-01 & 0.056 & 0.082 & 0.117 & 0.017 & 0.095 \\
                27   & LG/2\_2 & Sketches & 1.00e-02 & 0.061 & 0.08 & 0.105 & 0.022 & 2.761 \\
                28    & LG/2\_2 & Lines/n=10             & 1.00e-02 & 0.04  & 0.049 & 0.153 & 0.028 & 0.352\\
                29    & LG/2\_2 & Lines/n=20             & 1.00e-02 & 0.058 & 0.074 & 0.193 & 0.037 & 0.156\\
                30    & LG/2\_2 & Lines/n=50             & 1.00e-02 & 0.096 & 0.119 & 0.213 & 0.059 & 0.102\\
                31    & LG/2\_2 & Lines/n=100            & 1.00e-02 & 0.158 & 0.171 & 0.216 & 0.13  & 0.103\\
                32    & LG/2\_2 & Sketches/speckle noise & 2.00e+00 & 0.085 & 0.155 & 0.128 & 0.026 & 0.309\\
                33    & LG/2\_2 & Sketches/speckle noise & 1.00e+00 & 0.078 & 0.128 & 0.129 & 0.043 & 0.896\\
                34    & LG/2\_2 & Sketches/speckle noise & 1.00e-01 & 0.075 & 0.084 & 0.11  & 0.057 & 2.871\\
                35    & USTC   & Sketches               & 1.00e-02 & 0.086 & 0.114 & 0.16  & 0.135 & 0.187\\
        \end{tabular}
        \caption{\centering Summary of Synthetic Data Experiments. The training set consists of 1000 images and the test set consists of 100 images.}
        \label{tab:performance}
    \end{table*}

\hfill    
\normalem

\end{document}